\documentclass[12pt]{amsart}
\usepackage{amsmath,amssymb,color}
\newtheorem{corollary}{Corollary}[section]
\newtheorem{proposition}{Proposition}[section]

\newcommand{\<}{\langle}
\renewcommand{\>}{\rangle}
\newcommand{\uk}{|\kern-3.3pt\uparrow\>}
\newcommand{\dk}{|\kern-3.3pt\downarrow\>}
\newcommand{\ub}{\<\uparrow\kern-3.3pt|}
\newcommand{\db}{\<\downarrow\kern-3.3pt|}
\newcommand{\bs}{\boldsymbol}

\newcommand{\p}{\partial}
\newcommand{\re}{\text{Re}}
\newcommand{\im}{\text{Im}}

\newcommand{\ds}{\displaystyle}
\newcommand{\ts}{\textstyle}

\title[Matrix valued polynomials and quantum random walks]{Matrix valued Szeg\H o polynomials and quantum random walks}
\author[CGMV]{M. J. Cantero, F. A. Gr\"unbaum, L. Moral,  L. Vel\'{a}zquez}
\thanks{We thank Julia Kempe for some guidance with the literature.}
\thanks{The research of the second author was supported in part by NSF Grant \# DMS 0603901.}
\thanks{The research of the rest of the authors was partly supported by the Spanish grants from
the Ministry of Education and Science, project code MTM2005-08648-C02-01, and
the Ministry of Science and Innovation, project code MTM2008-06689-C02-01,
and by Project E-64 of Diputaci\'on General de Arag\'on (Spain).}
\date{}
\address[F. A. Gr\"unbaum]{Department of Mathematics \\ University of California \\ Berkeley \\ CA \\ 94720}
\address[M. J. Cantero, L. Moral, L. Vel\'{a}zquez]{Departamento de Matem\'{a}tica Aplicada \\ Universidad de Zaragoza \\ Zaragoza \\ Spain}

\begin{document}

\begin{abstract}
We consider quantum random walks (QRW) on the integers, a subject that has been
considered in the last few years in the framework of quantum computation.

We show how the theory of CMV matrices gives a natural tool to study these
processes and to give results that are analogous to those that Karlin and
McGregor developed to study (classical) birth-and-death processes using
orthogonal polynomials on the real line.

In perfect analogy with the classical case the study of QRWs on the set of
non-negative integers can be handled using scalar valued (Laurent)
polynomials and a scalar valued measure on the circle. In the case of
classical or quantum random walks on the integers one needs to allow for
matrix valued versions of these notions.

We  show how our tools yield results in the well known case of the
Hadamard walk, but we go beyond this translation invariant model to
analyze examples that are hard to analyze using other methods. More
precisely we consider QRWs on the set of non-negative integers. The
analysis of these cases leads to phenomena that are absent in the case of
QRWs on the integers even if one restricts oneself to a constant coin.
This is illustrated here by studying recurrence properties of the walk,
but the same method can be used for other purposes.

The presentation here aims at being selfcontained, but we refrain
from trying to give an introduction to quantum random walks, a
subject well surveyed in the literature we quote. For two excellent
reviews, see \cite{A,K}. See also the recent notes \cite{Ko}.
\end{abstract}

\maketitle

\section{Introduction and contents of the paper} \label{INT}

We start with a brief look at classical random walks.

Consider, for simplicity, a discrete time random walk in a denumerable
state space which we take to be the non-negative integers $i=0,1,2,\dots$.

The state of the system at time $n$ is given by a row vector $\pi_{i,n}$.
This is also called the probability distribution at time $n$. The $i^{th}$
component is interpreted as the probability that a particle can be found
at site $i$ at time $n$.

These non-negative quantities $\pi_{i,n}$ are assumed to add up to one,
for any $n$, when summed over $i$ in the set of non-negative integers.
This is not the only way to describe these simple processes, but this
facilitates the transition to the quantum case.

The evolution of the system is given by some transition probability matrix
$P=(P_{i,j})$. This means that for each state $i$ we have a collection of
``transition probabilities'' $P_{i,j}$ with the condition that these
non-negative numbers add up to one when summed over the index $j$.
$P_{i,j}$ is interpreted as the probability of a transition form site $i$
to site $j$ in one unit of time.

The Markov nature of our process is given by the fact that at time $t=1$
the new probability distribution gives weight
\[
\sum_{j=0}^{\infty}  \pi_{j,0}  P_{j,i}
\]
to the $i^{th}$ site. In other words the state at time $t=1$ is the vector
obtained by multiplying the probability distribution at time $0$, namely
$\pi_{j,0}$, by $P$.
The Markov property says that the process starts from scratch at any
integer time, so that the probability distribution (i.e., the state of the
system) at time $t=n$ is obtained by taking the product of the row vector
$\pi_{j,0}$ and the matrix $P^n$.

Random walks with a stationary (i.e., time invariant) transition mechanism
as above have been studied extensively. In some cases it is convenient to
exploit some of the mathematics associated with the matrix $P$. We will
not go into details here but it turns out that this is the case when the
matrix $P$ is either symmetric to begin with, i.e., $P_{i,j}=P_{j,i}$ or
it is what is called symmetrizable, namely there exists a vector $\pi_i$
such that
\[
\pi_i  P_{i,j} = P_{j,i} \pi_j.
\]

This notion is on the one hand related to the issue of reversibility and
on the other it allows one to introduce a certain inner product in
$L^2({\mathbb Z} \ge0$) that makes $P$ selfadjoint and thus one has
recourse to a nice and simple spectral theory.

A new item comes in now: the idea of a ``local'' transition. More
specifically we consider a special kind of random walks, namely those that
only allow for ``nearest neighbour'' transitions, i.e., $P_{i,j}=0$ if the
indices $i,j$ differ by more than one unit. In this case the matrix $P$ is
tridiagonal and this is the best of all worlds since under very simple
conditions any such matrix is symmetrizable into a tridiagonal one. Here
we see a physically important issue such as nearest neighbour interactions
going along with a nice mathematical fact such as symmetrizability.

The consequence of the combination of these two features is that one has a
powerful and natural tool to study random walks with a tridiagonal or
Jacobi transition matrix, namely the classical theory of orthogonal
polynomials on the real line, a subject that goes back to the beginning of
the 20th century.

This idea of making explicit use of the spectral theory for selfadjoint
operators in Hilbert space to study an important class of random walks on
${\mathbb Z} \ge 0$ probably appears first in \cite{KMc} . The authors
point out that similar ideas had been used earlier in the case of
diffusion processes by W. Feller, see \cite{F}, and H. P. McKean, Jr., see
\cite{McK}. One could add to this list of precursors of this fruitful line
of work other papers, such as \cite{LR}. One can also handle the case of
continuous time, but we do not pursue this here.

The basic idea in  \cite{KMc} is simple: if one is dealing with these
Markov chains that only allow for a ``local'' transition with $P_{i,j}=0$
if $i,j$ differ by more than one unit (also known as birth-and-death
processes) then the orthogonal polynomials that can be built out of the
corresponding Jacobi matrix and the underlying orthogonality measure give
all the ingredients of the spectral resolution of the matrix. In
particular they allow for a simple way to compute its power $P^n$.

All of this can be adapted to the case when our state space is not the set
of non-negative integers but the set of all integers. The fact that now we
have two different ways of going to infinity calls for the use of matrix
valued orthogonal polynomials, a notion introduced by M. G. Krein
\cite{K1,K2} around 1950. In this case we need polynomials with values in
the set of square matrices of size two, whereas up to now we were dealing
with scalar valued polynomials.

In fact a larger class of random walks, known under the name of
quasi-birth-and-death processes, can also be analyzed by using these
matrix valued orthogonal polynomials on the real line. In many cases the
orthogonality measure is not really a matrix valued measure, but an
appropriate linear functional. For two papers doing just this see
\cite{DRSZ,G1}.

With this background we can now move closer to the subject of our paper,
namely the study of a different sort of processes that go under the name
of ``quantum random walks'' (QRW).

The basic idea is this: a system has a (denumerable) set
$\{|i\>\}_{i\in\mathcal{I}}$ of measurable states called ``pure states".
The state $|\Psi_n\>$ of the system at time $n$ is a (complex)
superposition $|\Psi_n\>=\sum_{i\in\mathcal{I}}\psi_{i,n}|i\>$ of pure
states, so it is described by a ``wave function'' $\psi_{i,n}$ with $i$
running over the pure states. Therefore, a state can be identified with
its wave function.

The complex number $\psi_{i,n}$ is no longer a probability but a
probability amplitude, so that the actual probability to be in the state
$|i\>$ at time $n$ is $|\psi_{i,n}|^2$. The total probability of finding
the system at time $n$ must be 1, thus the wave function must satisfy the
condition
\begin{equation} \label{WF-NORM}
\sum_{i \in \mathcal{I}} |\psi_{i,n}|^2 = 1.
\end{equation}
In other words, the wave function lives in the unit ball of
$L^2(\mathcal{I})$, which is equivalent to consider the pure states as an
orthonormal basis of a Hilbert space whose unit vectors are the states of
the system.

According to the conservation of the probability, in the time invariant
case the time evolution of the system is characterized by a single unitary
operator $\mathfrak{U}$ on the Hilbert state space:
$|\Psi_n\>=\mathfrak{U}^n|\Psi_0\>$. If $U=(U_{i,j})_{i,j\in\mathcal{I}}$
is the unitary matrix such that
$\mathfrak{U}|i\>=\sum_{j\in\mathcal{I}}U_{i,j}|j\>$, the evolution of the
wave function is governed by
\begin{equation} \label{WF-EVOLUTION}
\psi_n = \psi_0U^n, \qquad \psi_n=(\psi_{i,n})_{i \in \mathcal{I}}.
\end{equation}
That is, the evolution operator in the wave function representation is a
unitary operator $U$ on $L^2(\mathcal{I})$.

When $\mathcal{I}=\mathbb{Z}$, it is proved in \cite{M} that all the local
QRWs ($U_{i,j}=0$ if $|i-j|>r$ for some $r$) which are translation
invariant ($U_{i,j}=U_{i+1,j+1}$) are, up to a phase, integer powers of a
simple translation, i.e., $\mathfrak{U}=e^{i\theta}T^k$, $T|j\>=|j+1\>$.

If we do not ask for translation invariance but allow only nearest
neighbour transitions ($U_{i,j}=0$ if $|i-j|>1$) then the QRW splits into
independent QRWs with no more than two pure states or, up to a change of
phases of the pure states, it is a simple right or left translation, i.e.,
$\mathfrak{U}|j\>=e^{i\theta_j}|j\pm1\>$. The situation is even worse in
the case $\mathcal{I}=\mathbb{Z}\ge0$, where the restriction to nearest
neighbour transitions always forces the splitting. These results, less
known in quantum computation, are direct consequences of \cite[Lemma
3.1]{BHJ} and \cite[Theorem 3.9]{MIN}.

A way to generate non-trivial QRWs on the integers or the non-negative
integers, allowing at the same time only nearest neighbour transitions, is
to include extra degrees of freedom. Following the analogy with a quantum
physical system, the simplest way to do this is to consider that every
site $i\in \mathcal{I}$ has an internal degree of freedom of spin 1/2
type, i.e, taking values $s\in\mathcal{S}=\{\uparrow,\downarrow\}$. The
space state is a tensor product spanned by
$\{|i\>\otimes\uk,|i\>\otimes\dk\}_{i\in\mathcal{I}}$.

We will be considering mostly the cases when $\mathcal{I}$ is either the
set of non-negative integers or the set of all integers. The evolution of
the wave function $\psi_{i,s,n}$ satisfies (\ref{WF-NORM}) when summing in
$i\in\mathcal{I}$ and $s\in\mathcal{S}$, so its evolution is given
similarly to (\ref{WF-EVOLUTION}) by a unitary operator $U$ on
$L^2(\mathcal{I}\times\mathcal{S})$.

Although the choice $\mathcal{I}=\mathbb{Z}\ge0$ would be as natural as
$\mathcal{I}=\mathbb{Z}$, the first one has not received too much
attention in the quantum case, maybe due to the difficulties to work with
a non translation invariant system. However, just as in the classical
case, QRWs on the non-negative integers are more natural for an orthogonal
polynomial approach. Indeed, from this point of view, QRWs on the
non-negative integers will be the cornerstone for the analysis of QRWs on
the integers.

We are finally ready to state the purpose of this paper. We will show that
in the case of a large class of quantum random walks on the integers there
is a natural tool that takes the place of the matrix valued orthogonal
polynomials on the real line briefly alluded to above, namely the theory of
Laurent $2\times 2$ matrix valued orthogonal polynomials associated with a
certain kind of unitary matrices, namely CMV matrices. This is a ready
made tool that combines the necessary features for a quantum mechanical
description of the phenomenon of nearest neighbours transitions: unitarity
and a block tridiagonal shape.

In the next section we first review the standard tools to deal with a
simple random walk on the integers: the first two are very well known and
have been used in the case of quatum random walks. The third method,
involving  matrix valued polynomials on the real line is less well known
as a useful tool for the study of classical random walks on the integers.
To the best of our knowledge neither scalar nor matrix valued orthogonal
polynomials have so far been used in the case of QRWs and they constitute
the main novel point of this paper.

The contents of the paper are organized as follows:

Section 2 reviews different approaches to the study of classical random
walks, with special emphasis in the methods using orthogonal polynomials.
This will establish a benchmark
for the development of analogous techniques in the quantum case.

Section 3 introduces CMV matrices and Szeg\H o polynomials.

Sections 4, 5 and 6 introduce QRWs and consider two extremely simple
examples to show the workings of our method.

Section 7 considers the case of a not (necessarily) constant coin. We
refer to this as dealing with ``distinct coins". In principle our method
can handle this general case, something that the more standard methods
cannot do.

Section 8 takes up the simpler case of a constant coin and one sees that
the case of the integers reduces to the study of two QRWs on the
non-negative integers.

Section 9 tackles in detail the case of a QRW with a constant coin on the
non-negative integers. The study of this case will be used in the
following section and it is of independent interest since the phases of
the coin have now a strong influence on the results, and in general the
orthogonality measure has a discrete mass.

Section 10 uses the results of the previous two sections and shows that
when dealing with all the integers there is no mass point in the orthogonality measure and that the
phases of the coin play no role.

Section 11 applies the results of the previous sections to compare in the
case of two specific QRWs their behaviour when considered on
$\mathbb{Z}\ge0$ or $\mathbb{Z}.$

Section 12 is devoted to getting some large time asymptotics.

Section 13 takes up the issue of recurrence of a QRW. By using our method
we see here some marked differences between the classical and the quantum
case. This section exhibits very clearly the benefits of using the notions
introduced in this paper to analyze QRWs.

Section 14 lists some conclusions as well as some open problems that can
be treated with the methods in this paper.

\section{Classical random walk on the integers, three looks at a classical subject} \label{CRW}

Consider a random walk on the integers with $p,q$ respectively the
probabilities of going right or left in one unit of time starting at any
integer position $i$.

Denote by $P(a,b,n)$ the quantity of interest, namely the
probability of going in $n$ steps from the initial position $a$ to
the final position $b$.

We describe three ways to study this basic problem.

\begin{itemize}

\item[a)] Path counting: we must have $c$ steps to the right and $d$ steps
to the left with $c+d=n$ and $c-d=b-a$. One can thus solve for $c,d$ in
terms of $a-b,n$. Each such path has probability $p^c q^d$ and there is
total of $n \choose c$ such paths. This gives $P(a,b,n)$.

\item[b)] Fourier methods: if for simplicity we take $p=q=1/2$,
it is clear that
\[
P(a,b,n)= \frac{1}{2} ( P(a,b-1,n-1) + P(a,b+1,n-1) ).
\]
Introduce the Fourier series
\[
\widehat{P}_a^r(\theta) \equiv \sum_{b=-\infty}^{\infty}
P(a,b,r)e^{ib\theta}
\]
with an inverse given by
\[
P(a,b,n) = \frac {1}{2\pi} \int_0^{2\pi} e^{-ib\theta} \widehat{P}_a^n(\theta)d\theta.
\]

The difference equation above can be used to see that
\[
\widehat{P}_a^n(\theta) = \widehat{P}_a^{n-1}(\theta)\cos \theta =
\widehat{P}_a^{0}(\theta)(\cos\theta)^n
\]
and therefore
\[
P(a,b,n) = \frac {1}{2\pi} \int_0^{2\pi}
\widehat{P}_a^0(\theta)(\cos\theta)^ne^{-ib\theta}d\theta.
\]
In the special case when $P(i,j,0)$ is given by $\delta_{i,j}$ we get
\[
P(a,b,n) = \frac {1}{2\pi} \int_0^{2\pi}
(\cos\theta)^ne^{-i(b-a)\theta}d\theta.
\]

This integral can be computed explicitly in terms of Bessel functions, or
just as easily, it can be seen to agree with the general expression
obtained earlier. For general values of $p,q$ one simply replaces
$\cos\theta$ by a weighted sum of $e^{i\theta},e^{-i\theta}$.

\medskip

These two methods have been properly adapted to study QRWs on the
integers. ``Path counting" was used by D. Meyer, see \cite{M}. This
analysis was pushed further along by using Jacobi polynomials in
\cite{ABNVW}. In this same paper as well as in \cite{NV} one finds
expressions obtained by using ``Fourier methods" to analyze the
appropriate recursion relations. The reader can find a very sophisticated
use of these formulas to derive some aymptotic results about the Hadamard
QRW in \cite{CIR}. A recent and very precise analysis of asymptotic
results is given in \cite{Ko}. Some interesting use of generating
functions to obtain rigorous results is made in \cite{BP}.

\medskip

\item[c)] Using matrix valued orthogonal polynomials:
since this is not too standard and since this is the way in which we
analyze the quantum case we give a lengthier  account of this way of
tackling this problem. This method goes back, in spirit at least, to the
original paper of Karlin and McGregor, \cite{KMc}, where the reader will
find a complete discussion of the case of a random walk on the
non-negative integers using scalar valued orthogonal polynomials. For a
fuller account of the material below see \cite{DRSZ,G1,G2}.

\end{itemize}

Consider, with M.~G.~Krein, \cite{K1,K2}, the set of polynomials of the
real variable $x$ with matrix coefficients of a fixed size $d$. All the
matrices that appear below have this common size $d$ and this is the
appropriate choice when the state space is the cartesian product of the
set of non-negative integers with the set ${1, 2, 3,\dots, d}$.

Given a positive definite matrix valued measure $dM(x)$ with finite
moments $\int_\mathbb{R}x^ndM(x)$, $n=0,1,2,\dots$, consider the skew
symmetric bilinear form defined for any pair of matrix valued polynomial
functions $P(x)$ and $Q(x)$ by the numerical matrix
\[
(P,Q)=(P,Q)_M = \int_{\mathbb R} P(x)dM(x)Q(x)^\dag,
\]
where $Q(x)^\dag$ denotes the conjugate transpose of $Q(x)$.

By the usual Gram--Schmidt construction this leads to the existence of a
sequence of matrix valued orthogonal polynomials $P_n(x) = K_{n,n} x^n +
K_{n,n-1} x^{n-1} + \cdots$ with non-singular leading coefficient
$K_{n,n}$. We make no special assumption on $K_{n,n}$.

Given an orthogonal sequence $\{P_n (x)\}_{n\ge 0}$ of matrix valued
orthogonal polynomials  one gets by the usual argument a three term
recursion relation
\begin{equation}\label{eq1}
xP_n(x) = A_{n}P_{n-1}(x) + B_nP_n(x) + C_{n}P_{n+1}(x),
\end{equation}
where $A_{n}$, $B_{n}$ and $C_{n}$ are matrices and the last one is
non-singular. If we had insisted on orthonormal polynomials then we would
get some relations among these coefficients.

It is convenient to introduce the block tridiagonal matrix $\mathbf{P}$
\[
\mathbf{P} = \begin{pmatrix}
B_0 & C_0  \\
A_1 & B_1 & C_1  \\
&\ddots &\ddots &\ddots
\end{pmatrix}.
\]

If $\mathbf{P}_{i,j}$ denotes the $i,j$ block of $\mathbf{P}$ we can
generate a sequence of $d \times d$ matrix valued polynomials $Q_i(x)$ by
imposing the three term recursion given above.  By using the notation of
the scalar case, we would have
\[
\mathbf{P} Q(x) = x Q(x)
\]
where the entries of the column vector $Q(x)$ are now $d \times d$
matrices.

Proceeding as in the scalar case, this relation can be iterated to give
\[
\mathbf{P}^n Q(x) = x^n Q(x)
\]
and if we assume the existence of a positive definite matrix valued measure
$dM(x)$ as in Krein's theory, with the property
\[
(Q_j,Q_j)\delta_{i,j} = \int_{\mathbb R} Q_i(x)dM(x)Q_j(x)^\dag,
\]
it is then clear that one can get an expression for the $i,j$ entry of the
block matrix $\mathbf{P}^n$ that would look exactly as in the scalar case,
namely
\[
(\mathbf{P}^n)_{i,j} (Q_j,Q_j) = \int_{\mathbb R} x^n Q_i(x)dM(x)Q_j(x)^\dag.
\]

The expression above, allowing one to compute the entries of
$\mathbf{P}^n$ is usually called the Karlin-McGregor formula, see
\cite{KMc}.

It may be worth noticing that the integrals above are different from the
ones that appear when one uses the Fourier method. The same will be true
in the quantum case.

Just as in the scalar case, the expression above becomes useful when
we can get our hands on the matrix valued polynomials $Q_i(x)$ and
the orthogonality measure $dM(x)$. When this is the case this
formula allows us to compute the transition probabilities between
any pair of states $i\le j$ {\bf in any number of steps} by using
only the top $j$ rows of the matrix $\mathbf{P}$. The ``time
dependence" has now been isolated to the term $x^n$ and the study of
its behaviour for large values of $n$ can be handled with
traditional methods. The same remark will apply in the quantum case
and this will be used in Section \ref{ASY}.

We are now ready to tackle the example of random walk on the integers,
when the probabilities of going right or left are $p$ and $q$
respectively. This is the most general translation invariant random walk
on the integers with nearest neighbours transitions. We present it here to
compare this analysis with its quantum counterpart, i.e., the quantum
random walks with a constant coin on the integers analyzed in Sections
\ref{CON} and \ref{CON-LINE}.

If we ``fold'' the integers by relabelling the natural sequence
\[
\dots -3,-2,-1,0,1,2,3\dots
\]
in the fashion
\[
\dots 5, 3, 1, 0, 2 , 4, 6 \dots
\]
then the transition probability matrix goes from being a scalar
tridiagonal doubly infinite one with $p$ in the ${i,i+1}$ diagonal and $q$
in the ${i+1,i}$ diagonal to the following semi-infinite block tridiagonal
matrix (with $2 \times 2$ blocks)
\[
\mathbf{P} = \begin{pmatrix}
0 & q & p & 0 & 0 & 0 & 0 & 0 & \dots \\
p & 0 & 0 & q & 0 & 0 & 0 & 0 & \dots \\
q & 0 & 0 & 0 & p & 0 & 0 & 0 & \dots \\
0 & p & 0 & 0 & 0 & q & 0 & 0 & \dots \\
0 & 0 & q & 0 & 0 & 0 & p & 0 & \dots \\
0 & 0 & 0 & p & 0 & 0 & 0 & q & \dots \\
\dots & \dots & \dots & \dots & \dots & \dots & \dots & \dots & \dots
\end{pmatrix},
\qquad
p+q = 1.
\]

For this first example the appropriate matrix measure is already found in
the original paper by S.~Karlin and J.~McGregor. It is given by
\[
dM(x) =
\frac {1}{\sqrt{4pq-x^2}}
\begin{pmatrix}
1 & x/2q \\
x/2q & p/q
\end{pmatrix}
dx,
\qquad
|x| \le \sqrt{4pq}.
\]

One does not find in the paper mentioned above the corresponding matrix
valued orthogonal polynomials or the block tridiagonal matrix, but they
can be easily given, see \cite{G1} and below.

In this example we have
\[
\begin{array}{c}
B_0 = \begin{pmatrix}0&q\\p&0\end{pmatrix},
\qquad
B_k= \begin{pmatrix}0&0\\0&0\end{pmatrix}, \quad k\geq 1,
\bigskip \\
A_k=\begin{pmatrix}q&0\\0&p\end{pmatrix}, \quad k\geq 1,
\qquad\quad
C_k=\begin{pmatrix}p&0\\0&q\end{pmatrix}, \quad k\geq 0.
\end{array}
\]
The orthogonal polynomials given by
\[
\begin{array}{l}
A_kP_{k-1}(x)+B_kP_k(x)+C_kP_{k+1}(x)=xP_k(x),
\medskip \\
P_{-1}(x)=\bs{0}, \quad P_0(x)=\bs{1},
\end{array}
\]
where $\bs{0}$ and $\bs{1}$ are the null and identity matrix, can be
easily expressed in terms of Chebyshev polynomials.

Let us denote by $U_n(x)$ the Chebyshev polynomials of the second kind,
which satisfy
\begin{equation} \label{eq2}
U_{n+1}(x)+U_{n-1}(x)=2xU_n(x), \qquad U_{-1}(x)=0 \qquad U_0(x)=1.
\end{equation}
The relation with the Chebyshev polynomials $U_n(x)$, is given by
\[
P_k(x) = \begin{pmatrix} (q/p)^\frac{k}{2} & \kern-10pt 0 \\ 0 & \kern-10pt (p/q)^\frac{k}{2} \end{pmatrix}
\left\{
\bs{1}U_k(x^{*}) -
\begin{pmatrix} 0 & \kern-5pt (q/p)^\frac{1}{2} \\ (p/q)^\frac{1}{2} & \kern-5pt 0 \end{pmatrix} U_{k-1}(x^{*})
\right\},
\]
where $x^{*}=x/2\sqrt{pq}$.
This expression can be compared to the one in Section \ref{CON-LINE}.

These ideas have been used recently to analyze more elaborate examples
of classical random walks. The reader may consult \cite{DRSZ,G1,G2,G3,GI}.

\section{Szeg\H o polynomials and CMV matrices} \label{CMV}

All the information about a QRW is encoded in the unitary operator $U$
governing the evolution of the system. Therefore, it is not strange that
the theory of canonical matrix representations of unitary operators on
Hilbert spaces should play an important role in the study of QRWs.
Surprisingly, such a theory has been developed only recently (see
\cite{FIVE,MIN,Wa93}) giving rise to the so called CMV matrices, related
to the Szeg\H o polynomials.

Due to their relevance for the rest of the paper, we will summarize in
this section the main facts about CMV matrices.

The basic idea is that, as a consequence of the spectral theorem, any
unitary operator is unitarily equivalent to a direct sum of unitary
multiplication operators, i.e., operators of the type
\begin{equation} \label{UMO}
U_\mu \colon \mathop{L^2_\mu(\mathbb{T}) \to L^2_\mu(\mathbb{T})}
\limits_{\ds \, f(z) \longrightarrow zf(z)}
\end{equation}
$\mu$ being a probability measure on the unit circle
$\mathbb{T}=\{z\in\mathbb{C}:|z|=1\}$, and $L^2_\mu(\mathbb{T})$ the
Hilbert space of $\mu$-square-integrable functions with inner product
\[
(f,g) = \int_\mathbb{T} \overline{f(z)}\,g(z)\,d\mu(z).
\]
Thus, it is enough to discuss the canonical representations of unitary
multiplication operators. Moreover, we can suppose that $\mu$ has an
infinite support, otherwise $L^2_\mu(\mathbb{T})$ is finite-dimensional,
so $U_\mu$ is unitarily diagonalizable.

Since the Laurent polynomials are dense in $L^2_\mu(\mathbb{T})$, a
natural basis to obtain a matrix representation of $U_\mu$ is given by the
Laurent polynomials $(\chi_j)_{j=0}^\infty$ obtained from the Gram-Schmidt
orthonormalizalization of $\{1,z,z^{-1},z^2,z^{-2},\dots\}$ in
$L^2_\mu(\mathbb{T})$.

The matrix $\mathcal{C}=(\chi_j,z\chi_k)_{j,k=0}^\infty$ of $U_\mu$ with
respect to $(\chi_j)_{j=0}^\infty$ has the form
\begin{equation} \label{C}
\mathcal{C} =
\begin{pmatrix}
\overline{\alpha}_0 & \kern-3pt \rho_0\overline{\alpha}_1 & \kern-3pt \rho_0\rho_1
& \kern-3pt 0 & \kern-3pt 0 & \kern-3pt 0 & \kern-3pt 0 & \kern-3pt \dots
\\
\rho_0 & \kern-3pt -\alpha_0\overline{\alpha}_1 & \kern-3pt -\alpha_0\rho_1
& \kern-3pt 0 & \kern-3pt 0 & \kern-3pt 0 & \kern-3pt 0 & \kern-3pt \dots
\\
0 &
\kern-3pt \rho_1\overline{\alpha}_2 & \kern-3pt -\alpha_1\overline{\alpha}_2 & \kern-3pt \rho_2\overline{\alpha}_3 & \kern-3pt \rho_2\rho_3
& \kern-3pt 0 & \kern-3pt 0 & \kern-3pt \dots
\\
0 &
\kern-3pt \rho_1\rho_2 &  \kern-3pt -\alpha_1\rho_2 & \kern-3pt -\alpha_2\overline{\alpha}_3 & \kern-3pt -\alpha_2\rho_3
& \kern-3pt 0 & \kern-3pt 0 & \kern-3pt \dots
\\
0 & \kern-3pt 0 &
\kern-3pt 0 & \kern-3pt \rho_3\overline{\alpha}_4 & \kern-3pt -\alpha_3\overline{\alpha}_4 & \kern-3pt \rho_4\overline{\alpha}_5 & \kern-3pt \rho_4\rho_5
& \kern-3pt \dots
\\
0 & \kern-3pt 0 &
\kern-3pt 0 & \kern-3pt \rho_3\rho_4 & \kern-3pt -\alpha_3\rho_4 & \kern-3pt -\alpha_4\overline{\alpha}_5 & \kern-3pt -\alpha_4\rho_5
& \kern-3pt \dots
\\
\dots & \kern-3pt \dots & \kern-3pt \dots & \kern-3pt \dots & \kern-3pt \dots & \kern-3pt \dots & \kern-3pt \dots & \kern-3pt \dots
\end{pmatrix},
\end{equation}
where $\rho_j=\sqrt{1-|\alpha_j|^2}$ and $(\alpha_j)_{j=0}^\infty$ is a
sequence of complex numbers such that $|\alpha_j|<1$. The coefficients
$\alpha_j$ are known as the Verblunsky parameters of the measure $\mu$,
and establish a bijection between the probability measures supported on an
infinite set of the unit circle and the sequences in the open unit disk.

Another equally natural basis would be the Laurent polynomials
$(x_j)_{j=0}^\infty$ obtained from the orthonormalization of
$\{1,z^{-1},z, z^{-2},z^2,\dots\}$. They are given by
\[
x_j(z)=\overline{\chi_j(1/\overline z)}
\]
and, consequently, the matrix of $U_\mu$ with respect to
$(x_j)_{j=0}^\infty$ is the transpose $\mathcal{C}^T$ of $\mathcal{C}$.

As a consequence, we have the identities
\begin{equation} \label{CX}
\begin{aligned}
\chi(z)\mathcal{C}&=z\chi(z), \qquad \chi = (\chi_0,\chi_1,\chi_2,\dots), \qquad \chi_0=1,
\\
\mathcal{C}x(z)&=zx(z), \qquad x = (x_0,x_1,x_2,\dots)^T, \qquad x_0=1,
\end{aligned}
\end{equation}
which can be viewed as recurrences which determine the orthonormal Laurent
polynomials.

The unitary matrices with the form (\ref{C}) or its transpose are called
CMV matrices, but we will reserve this name for the matrix in (\ref{C}).
Also, when talking about orthonormal Laurent polynomials, we will refer to
$(x_j)_{j=0}^\infty$, which are the ones that we will normally use.

The canonical representations of the unitaries are the narrowest
banded representations that can be obtained for all such operators.
The previous results state that every unitary operator has a matrix
representation which is a direct sum of CMV matrices, so the
canonical representations of the unitaries are at least
five-diagonal. That they are exactly five-diagonal has been proved
in \cite{MIN}, where it was shown that not every unitary operator
admits a four-diagonal representation.

The CMV matrices have also a tridiagonal factorization
$\mathcal{C}=\mathcal{L}\mathcal{M}$, with two unitary $2\times2$-block
diagonal symmetric factors given by
\begin{equation} \label{LM1}
\begin{aligned}
& \mathcal{L} = \text{diag}(\Theta_0,\Theta_2,\Theta_4,\dots),
\\
& \mathcal{M} = \text{diag}(1,\Theta_1,\Theta_3,\dots),
\end{aligned}
\qquad
\Theta_j = \begin{pmatrix} \overline{\alpha}_j & \rho_j \\ \rho_j & -\alpha_j \end{pmatrix}.
\end{equation}

The Verblunsky parameters have a special meaning in terms of the Szeg\H o
polynomials $(\varphi_j)_{j=0}^\infty$, which come from orthonormalizing
$\{z^j\}_{j=0}^\infty$, see \cite{Sz75,Ge61}.

These polynomials are not so useful as a basis because the polynomials are
not always dense in $L^2_\mu(\mathbb{T})$. Nevertheless, they are related
to the orthonormal Laurent polynomials by
\begin{equation} \label{OP-OLP}
\begin{aligned}
& \chi_{2j}(z) = z^{-j}\varphi_{2j}^*(z), \qquad \chi_{2j+1}(z) = z^{-j} \varphi_{2j+1}(z),
\\
& x_{2j}(z) = z^{-j}\varphi_{2j}(z), \qquad x_{2j+1}(z) = z^{-j-1} \varphi_{2j+1}^*(z),
\end{aligned}
\end{equation}
where $\varphi_j^*(z)=z^j\overline{\varphi_j(1/\overline z)}$.

The key result is that the Szeg\H o polynomials are determined by the
recurrence relation
\begin{equation} \label{RR-OP}
\rho_j\varphi_{j+1}(z) = z\varphi_j(z) - \overline{\alpha}_j \varphi_j^*(z), \qquad \varphi_0=1,
\end{equation}
so the recurrence for the monic orthogonal polynomials $(\phi_j)_{j=0}^\infty$ is
\begin{equation} \label{RR-MOP}
\phi_{j+1}(z) = z\phi_j(z) - \overline{\alpha}_j \phi_j^*(z), \qquad \phi_0=1,
\end{equation}
which shows that $\alpha_j=-\overline{\phi_{j+1}(0)}$.

For some measures on the unit circle the Szeg\H o polynomials, and
therefore the Verblunsky parameters and the CMV matrix, are known
explicitly, see \cite{Si04-1}. Other cases can be analyzed following
different methods.

For instance, given a sequence of Verblunsky parameters, the spectral
analysis of the corresponding CMV matrix can be used to obtain information
about the orthogonality measure because it is related to the spectral
measure of the CMV matrix. Indeed, the support of the measure coincides
with the spectrum of the CMV matrix, the mass points $z_0$ being the
eigenvalues, which are simple and have eigenvectors given by $x(z_0)$.
Bearing in mind (\ref{CX}), this means that $x(z_0) \in
L^2(\mathbb{Z}\ge0)$ exactly when $z_0$ is a mass point. Actually,
$\mu(\{z_0\})=1/\|x(z_0)\|^2$.

These results will be of interest later on, see Section \ref{ASY}.

Perturbative results are useful to study new examples taking as a starting
point known ones. The simplest example of this is a rotation of the
measure by an angle $\vartheta$, i.e.
\[
d\mu(z) \to d\mu(e^{-i\vartheta}z).
\]
The change of the monic orthogonal polynomials $\phi_j(z) \to
e^{ij\vartheta}\phi_j(e^{-i\vartheta}z)$ shows that the effect of the
rotation on the Verblunsky parameters is
\begin{equation} \label{ROT-VP}
\alpha_j \to e^{-i(j+1)\vartheta}\alpha_j
\end{equation}
while (\ref{OP-OLP}) implies that the orthonormal Laurent polynomials
transform as
\begin{equation} \label{ROT-OLP}
x_{2j-1}(z) \to e^{-ij\vartheta}x_{2j-1}(e^{-i\vartheta}z),
\qquad
x_{2j}(z) \to e^{ij\vartheta}x_{2j}(e^{-i\vartheta}z).
\end{equation}

The transformation above will play an important role in later sections.

Another tool for the study of Szeg\H o polynomials is the Carath\'{e}odory
function $F$ of the orthogonality measure $\mu$, defined by
\begin{equation} \label{CF}
F(z) = \int_\mathbb{T} \frac{t+z}{t-z}\,d\mu(t),
\qquad |z|<1.
\end{equation}
$F$ is analytic on the open unit disc with McLaurin series
\begin{equation} \label{MOM}
F(z) = 1 + 2\sum_{j=1}^\infty\overline\mu_jz^j,
\qquad
\mu_j = \int_\mathbb{T} z^jd\mu(z),
\end{equation}
whose coefficients provide the moments $\mu_j$ of the measure $\mu$.

$F$ can be obtained as
\begin{equation} \label{CF-OP}
F(z) = \lim_{j\to\infty} \frac{\tilde{\varphi}_j^*(z)}{\varphi_j^*(z)},
\qquad |z|<1,
\end{equation}
where $\tilde{\varphi}_j$ are the Szeg\H o polynomials whose Verblunsky
parameters are given by $-\alpha_j$ if the original ones were $\alpha_j$.

The Carath\'{e}odory function is a shortcut that allows one to recover the
measure from the Szeg\H o polynomials. If
\begin{equation} \label{RN}
d\mu(e^{i\theta})=w(\theta)\frac{d\theta}{2\pi}+d\mu_s(e^{i\theta}),
\qquad \theta\in(-\pi,\pi],
\qquad \mu_s \text{ singular},
\end{equation}
the weight $w(\theta)$ is given by
\begin{equation} \label{CF-w}
w(\theta) = \lim_{r\uparrow1} \re F(re^{i\theta}),
\end{equation}
and the support of $\mu_s$ lies on $\{e^{i\theta} :
\lim_{r\uparrow1}F(re^{i\theta}) = \infty \}$. In particular,
$e^{i\theta_0}$ is a mass point of $\mu$ with mass
$\mu(\{e^{i\theta_0}\})$ if and only if
\begin{equation} \label{MASS}
\mu(\{e^{i\theta_0}\}) = \lim_{r\uparrow1}\frac{1-r}{2}F(re^{i\theta_0}) \neq 0.
\end{equation}

Two natural extensions of CMV matrices are of interest to us: doubly
infinite CMV matrices and block CMV matrices.

Given a two-sided sequence $(\alpha_j)_{j\in\mathbb{Z}}$ we can define
the doubly infinite CMV matrix $\mathcal{C}=\mathcal{L}\mathcal{M}$, where
\[
\begin{aligned}
& \mathcal{L}=
\text{diag}(\dots,\Theta_{-4},\Theta_{-2},\Theta_0,\Theta_2,\Theta_4,\dots),
\\
& \mathcal{M}=
\text{diag}(\dots,\Theta_{-3},\Theta_{-1},\Theta_1,\Theta_3,\dots),
\end{aligned}
\]
and $\Theta_j$, given in (\ref{LM1}), acts on the indices $j$ and $j+1$.
We will use the same notations as for semi-infinite CMV matrices.

$\mathcal{C}$ is a five-diagonal doubly infinite unitary matrix with the
form
\[
\begin{array}{cccc|cccc}
\dots & \kern-3pt\dots & \kern-3pt\dots & \kern-3pt\dots
\kern-2pt & \kern-2pt
\dots & \kern-3pt\dots & \kern-3pt\dots & \kern-3pt\dots
\\
\dots & \kern-3pt \rho_{-3}\overline{\alpha}_{-2} & \kern-3pt -\alpha_{-3}\overline{\alpha}_{-2} & \kern-3pt \rho_{-2}\overline{\alpha}_{-1}
\kern-2pt & \kern-2pt
\rho_{-2}\rho_{-1} & \kern-3pt 0 & \kern-3pt 0 & \kern-3pt \dots
\\
\dots & \kern-3pt \rho_{-3}\rho_{-2} & \kern-3pt -\overline{\alpha}_{-3}\rho_{-2} & \kern-3pt -\alpha_{-2}\overline{\alpha}_{-1}
\kern-2pt & \kern-2pt
-\alpha_{-2}\rho_{-1} & \kern-3pt 0 & \kern-3pt 0 & \kern-3pt \dots
\\ \hline
\dots & \kern-3pt 0 & \kern-3pt 0 & \kern-3pt \rho_{-1}\overline{\alpha}_0
\kern-2pt & \kern-2pt
-\alpha_{-1}\overline{\alpha}_0 & \kern-3pt \rho_0\overline{\alpha}_1 & \kern-3pt \rho_0\rho_1 & \kern-3pt \dots
\\
\dots & \kern-3pt 0 & \kern-3pt 0 & \kern-3pt \rho_{-1}\rho_0 & -\alpha_{-1}\rho_0
\kern-2pt & \kern-2pt
\kern-3pt -\alpha_0\overline{\alpha}_1 & \kern-3pt -\alpha_0\rho_1 & \kern-3pt \dots
\\
\dots & \kern-3pt \dots & \kern-3pt \dots & \kern-3pt \dots
\kern-2pt & \kern-2pt
\dots & \kern-3pt \dots & \kern-3pt\dots & \kern-3pt\dots
\end{array}
\]

Doubly infinite CMV matrices are also related to the other generalization
of interest: block CMV matrices. They appear in connection with matrix
valued Szeg\H o polynomials, see \cite{DaPuSi08,Si04-1,SiFIVE}.

Given a positive definite $d \times d$ matrix valued measure $\bs{\mu}$ on
the unit circle we can define the right and left ``inner products"
\[
\bs{(f,g)}_L=\int_\mathbb{T} \bs{g}(z) \, d\bs{\mu}(z) \, \bs{f}(z)^\dag,
\qquad \bs{(f,g)}_R=\int_\mathbb{T} \bs{f}(z)^\dag d\bs{\mu}(z) \,
\bs{g}(z),
\]
for $\bs{f},\bs{g}$ with values in the set of $d \times d$ matrices.
Notice that the symbol $^\dag$ which denotes the adjoint matrix includes
the conjugation of $z$. In what follows we suppose that
$\int_\mathbb{T}d\bs{\mu}(z)=\bs{1}$ is the $d \times d$ unit matrix.

Now we can consider right and left matrix valued Szeg\H o polynomials,
$(\bs{\varphi}_j^R)_{j=0}^\infty$ and $(\bs{\varphi}_j^L)_{j=0}^\infty$,
arising from the standard orthonormalization of
$\{\bs{1}z^j\}_{j=0}^\infty$ with respect to $\bs{(,)}_R$ and $\bs{(,)}_L$
respectively.

Analogously to the scalar case, the matrix valued Szeg\H o polynomials
satisfy a recurrence given in terms of a sequence
$(\bs{\alpha}_j)_{j=0}^\infty$ of $d \times d$ matrices such that
$\|\bs{\alpha}_j\|<1$. However, this recurrence mixes the right and left
polynomials in the following way
\[
\begin{aligned}
& \bs{\rho}_j^L\bs{\varphi}_{j+1}^L(z) = z\bs{\varphi}_j^L(z) - \bs{\alpha}_j^\dagger\bs{\varphi}_j^{R*}(z),
\\
& \bs{\varphi}_{j+1}^R(z) \bs{\rho}_j^R = z\bs{\varphi}_j^R(z) - \bs{\varphi}_j^{L*}(z)\bs{\alpha}_j^\dagger,
\end{aligned}
\qquad \bs{\varphi}_0^L = \bs{\varphi}_0^R=\bs{1},
\]
where $\bs{\varphi}_j^*(z) = z^j \bs{\varphi}_j(1/\overline{z})^\dag$
and $\bs{\rho}^L_j,\bs{\rho}^R_j$ are the positive definite matrices
\[
\bs{\rho}_j^L=(1-\bs{\alpha}_j^\dag\bs{\alpha}_j)^{1/2},
\qquad
\bs{\rho}_j^R=(1-\bs{\alpha}_j\bs{\alpha}_j^\dag)^{1/2}.
\]

This mixture between right and left polynomials also appears in the
connection with the orthonormal Laurent polynomials
$(\bs{\chi}_j)_{j=0}^\infty,(\bs{x}_j)_{j=0}^\infty$ given by
\begin{equation} \label{OP-OLP-Matr}
\begin{aligned}
& \bs{\chi}_{2j}(z) = z^{-j}\bs{\varphi}_{2j}^{L*}(z), \qquad \bs{\chi}_{2j+1}(z) = z^{-j}\bs{\varphi}_{2j+1}^R(z),
\\
& \bs{x}_{2j}(z) = z^{-j}\bs{\varphi}_{2j}^L(z), \qquad \bs{x}_{2j+1}(z) = z^{-j-1} \bs{\varphi}_{2j+1}^{R*}(z),
\end{aligned}
\end{equation}
and related by $\bs{x}_j(z)=\bs{\chi}_j(1/\overline{z})^\dag$.
$(\bs{\chi}_j)_{j=0}^\infty$ comes from the orthonormalization of
$\{\bs{1},\bs{1}z,\bs{1}z^{-1},\bs{1}z^2,\bs{1}z^{-2},\dots\}$ with
respect to $\bs{(,)}_R$, and $(\bs{x}_j)_{j=0}^\infty$ from the
orthonormalization of
$\{\bs{1},\bs{1}z^{-1},\bs{1}z,\bs{1}z^{-2},\bs{1}z^2,\dots\}$ with
respect to $\bs{(,)}_L$ (notice a slight difference with respect to
\cite{DaPuSi08}, where the two kinds of Laurent polynomials discussed are
both orthonormal with respect to $\bs{(,)}_R$).

The matrix $\bs{\mathcal{C}}$ which determines the orthonormal Laurent
polynomials through the recurrences
\begin{equation} \label{CX-Matr}
\begin{aligned}
\bs{\chi}(z)\bs{\mathcal{C}}&=z\bs{\chi}(z),
\qquad \bs{\chi} = (\bs{\chi}_0,\bs{\chi}_1,\bs{\chi}_2,\dots), \qquad \bs{\chi}_0=\bs{1},
\\
\bs{\mathcal{C}}\bs{x}(z)&=z\bs{x}(z),
\qquad \bs{x} = (\bs{x}_0,\bs{x}_1,\bs{x}_2,\dots)^T, \qquad \bs{x}_0=\bs{1},
\end{aligned}
\end{equation}
is given by $\bs{\mathcal{C}}=\bs{\mathcal{L}\mathcal{M}}$ with
\begin{equation} \label{LM2}
\begin{aligned}
& \bs{\mathcal{L}} = \text{diag}(\bs{\Theta}_0,\bs{\Theta}_2,\bs{\Theta}_4,\dots),
\\
& \bs{\mathcal{M}} = \text{diag}(\bs{1},\bs{\Theta}_1,\bs{\Theta}_3,\dots),
\end{aligned}
\qquad
\bs{\Theta}_j = \begin{pmatrix} \bs{\alpha}_j^\dag & \bs{\rho}_j^L \\ \bs{\rho}_j^R & -\bs{\alpha}_j \end{pmatrix}.
\end{equation}

The unitary matrix $\bs{\mathcal{C}}$ is called a block CMV matrix. Its
explicit form is
\[
\bs{\mathcal{C}} = \begin{pmatrix}
\bs{\alpha}_0^\dagger & \kern-3pt \bs{\rho}_0^L\bs{\alpha}_1^\dagger & \kern-3pt \bs{\rho}_0^L\bs{\rho}_1^L &
\kern-3pt \bs{0} & \kern-3pt \bs{0} &  \kern-3pt \bs{0} & \kern-3pt \bs{0} & \kern-3pt\dots
\\
\bs{\rho}_0^R & \kern-3pt -\bs{\alpha}_0\bs{\alpha}_1^\dagger & \kern-3pt -\bs{\alpha}_0\bs{\rho}_1^L &
\kern-3pt \bs{0} & \kern-3pt \bs{0} & \kern-3pt \bs{0} & \kern-3pt \bs{0} & \kern-3pt\dots
\\
\bs{0} &
\kern-3pt \bs{\alpha}_2^\dagger\bs{\rho}_1^R & \kern-3pt -\bs{\alpha}_2^\dagger\bs{\alpha}_1 &
\kern-3pt \bs{\rho}_2^L\bs{\alpha}_3^\dagger & \kern-3pt \bs{\rho}_2^L\bs{\rho}_3^L &
\kern-3pt \bs{0} & \kern-3pt \bs{0} & \kern-3pt\dots
\\
\bs{0} &
\kern-3pt \bs{\rho}_2^R\bs{\rho}_1^R & \kern-3pt -\bs{\rho}_2^R\bs{\alpha}_1 &
\kern-3pt -\bs{\alpha}_2\bs{\alpha}_3^\dagger & \kern-3pt -\bs{\alpha}_2\bs{\rho}_3^L &
\kern-3pt \bs{0} & \kern-3pt \bs{0} & \kern-3pt\dots
\\
\bs{0} & \kern-3pt \bs{0} & \kern-3pt \bs{0} &
\kern-3pt \bs{\alpha}_4^\dagger\bs{\rho}_3^R & \kern-3pt -\bs{\alpha}_4^\dagger\bs{\alpha}_3 &
\kern-3pt \bs{\rho}_4^L\bs{\alpha}_5^\dagger & \kern-3pt \bs{\rho}_4^L \bs{\rho}_5^L &
\kern-3pt\dots
\\
\bs{0} & \kern-3pt \bs{0} & \kern-3pt \bs{0} &
\kern-3pt \bs{\rho}_4^R\bs{\rho}_3^R & \kern-3pt -\bs{\rho}_4^R\bs{\alpha}_3 &
\kern-3pt -\bs{\alpha}_4\bs{\alpha}_5^\dagger & \kern-3pt -\bs{\alpha}_4\bs{\rho}_5^L &
\kern-3pt\dots
\\
\dots & \kern-3pt\dots & \kern-3pt\dots & \kern-3pt\dots & \kern-3pt\dots & \kern-3pt\dots & \kern-3pt\dots & \kern-3pt\dots
\end{pmatrix},
\]
where the symbol $\bs{0}$ stands for the $d \times d$ null matrix.

Particularly simple is the case of diagonal Verblunsky parameters
$\bs{\alpha}_j=\text{diag}(\alpha^1_j,\dots,\alpha^d_j)$. The
corresponding matrix orthonormal polynomials and orthogonality matrix
measure are diagonal too. In other words, the recurrences in
(\ref{CX-Matr}) split into $d$ ones associated with scalar CMV matrices
with Verblunsky parameters $\alpha^1_j,\dots,\alpha^d_j$.

Finally, the matrix valued Carath\'{e}odory function
\begin{equation} \label{CF-Matr}
\bs{F}(z) = \int_\mathbb{T} \frac{t+z}{t-z}\,d\bs{\mu}(t),
\qquad |z|<1,
\end{equation}
and its real part $\re\bs{F}(z)=(\bs{F}(z)+\bs{F}(z)^\dag)/2$ allows one
to recover the matrix measure $\bs{\mu}$ just as in the scalar case. In
particular, the matrix moments $\bs{\mu}_j$ of $\bs{\mu}$ come from the
McLaurin series of $\bs{F}$, i.e.,
\begin{equation} \label{MOM-Matr}
\bs{F}(z) = \bs{1} + 2\sum\bs{\mu}_j^\dag z^j,
\qquad
\bs{\mu}_j = \int_\mathbb{T} z^j d\bs{\mu}(z).
\end{equation}

\section{Quantum random walks} \label{QRW}

We will consider a few one-dimensional quantum random walks with pure
states $|i\>\otimes\uk$ and $|i\>\otimes\dk$, where $i$ runs over the
non-negative integers or over all the integers, and with a one step
transition mechanism given by a unitary matrix $U$. Our goal here is to
give an explicit expression for the entries of the matrix $U^n$ for
$n=0,1,2,\dots$ describing the time evolution of our process. More
precisely, we will find a Karlin-McGregor (KMcG) formula for such entries.


The relevance of the CMV matrices, as the canonical representations of the
unitaries, is clear. However, the reduction of an arbitrary infinite
unitary matrix to CMV form can be a difficult task. Nevertheless, the QRWs
usually considered in the literature include only nearest neighbour
transitions. In this case it is possible to prove that the CMV form is
obtained after a simple change of phases in the basis, see \cite[Theorem
3.2 and Lemma 3.7]{MIN}. Then, the relation of the CMV matrices with the
Szeg\H o polynomials provides the functions and measures needed to obtain
a KMcG formula.

\section{A very elementary example} \label{FREE}

A simple QRW on the non-negative integers corresponds to the unitary
operator
\begin{equation} \label{S+}
S_+ = \sum_{i=0}^\infty|i+1\>\<i|\otimes\uk\ub
+ \sum_{i=1}^\infty|i-1\>\<i|\otimes\dk\db
+ |0\>\<0|\otimes\uk\db,
\end{equation}
which means that the transition mechanism moves spins at positions
$0,1,2,\dots$ according to the following (deterministic) prescription:
spins up at any location move one step to the right, spins down at
$1,2,3,\dots$ move to the left, and finally a spin down at location $0$
reverses orientation and stays at location $0$. In the next unit of time
this spin will move to location $1$.

If we choose to order the pure states of our system as follows
\begin{equation} \label{ORD}
|0\>\otimes\uk,|0\>\otimes\dk,|1\>\otimes\uk,|1\>\otimes\dk,\dots
\end{equation}
then the transition matrix is
\[
U_+ =
\begin{pmatrix}
0 & 0 & 1 & 0 & 0 & 0 & 0 & \dots \\
1 & 0 & 0 & 0 & 0 & 0 & 0 & \dots \\
0 & 0 & 0 & 0 & 1 & 0 & 0 & \dots \\
0 & 1 & 0 & 0 & 0 & 0 & 0 & \dots \\
0 & 0 & 0 & 0 & 0 & 0 & 1 & \dots \\
0 & 0 & 0 & 1 & 0 & 0 & 0 & \dots \\
\dots & \dots & \dots & \dots & \dots & \dots & \dots & \dots
\end{pmatrix}.
\]
We refer to the pure states, ordered in this fashion, as the zeroth, the
first, the second state, etc.

However, there is another way to look at the transition matrix. We could
think of the functions $1,z^{-1},z,z^{-2},z^2,\dots$ defined on the unit
circle and form a column vector $x(z)$ with these functions. Applying our
matrix $U_+$ to this vector is the same as multiplying the vector by $z$,
which is another way of saying that for each value of $z$ we have found a
formal eigenvector with eigenvalue $z$ for our matrix.

If we denote the components of this column vector by $x_j(z)$,
$j=0,1,2,\dots$, we claim that the probability amplitude of going in $n$
(positive or negative) units of time from a pure state $j$ to a pure state
$k$ (the indices $j,k$ run over all the zeroth, first, second, $\dots$
pure states introduced above) is given by the integral
\[
\frac{1}{2\pi i} \int_{|z|=1} \overline{x_k(z)}\,x_j(z)z^n \frac{dz}{z}.
\]
It is completely elementary to see that this integral is either $0$ or $1$
and that in fact we get the correct expression for the amplitudes.

The above relation is not an accident. In fact, this example is the
prototype of everything that follows. The matrix given above is the
simplest example of CMV matrix, corresponding to null Verblunsky
parameters. The Laurent polynomials $(x_j)_{j=0}^\infty$ are orthonormal
with respect to the Lebesgue measure $dz/2\pi iz=d\theta/2\pi$. Therefore,
the identity $Ux(z)=zx(z)$ gives
\[
\int_\mathbb{T} \overline{x_k(z)}\,z^nx_j(z) \frac{dz}{2\pi iz} =
\sum_{l=0}^\infty \int_\mathbb{T} \overline{x_k(z)}\,(U^n)_{j,l}\,x_l(z) \frac{dz}{2\pi iz} =
(U^n)_{j,k}.
\]

\section{Another example} \label{FREE2}

The natural extension of the previous example to the integers is the QRW
associated with the unitary operator
\begin{equation} \label{S}
S = \sum_{i\in\mathbb{Z}}|i+1\>\<i|\otimes\uk\ub +
\sum_{i\in\mathbb{Z}}|i-1\>\<i|\otimes\dk\db.
\end{equation}
The evolution is now even simpler than the previous one: all spins up move
to the right and all spins down move to the left.

Ordering the pure states as follows
\[
\dots,
|-1\>\otimes\uk,|-1\>\otimes\dk,
|0\>\otimes\uk,|0\>\otimes\dk,
|1\>\otimes\uk,|1\>\otimes\dk,
\dots
\]
the transition matrix $U$ is the doubly infinite matrix
\[
\begin{array}{ccccccc|ccccccc}
\ddots & \ddots & \ddots & \ddots & \ddots & & \\
& 0 & 0 & 0 & 0 & \kern3pt 1 & \\
& & 1 & 0 & 0 & \kern3pt 0 & \kern6pt 0 \kern3pt \\
& & & 0 & 0 & \kern3pt 0 & \kern6pt 0 \kern3pt & \kern3pt 1 \\
& & & & 1 & \kern3pt 0 & \kern6pt 0 \kern3pt & \kern3pt 0 & \kern6pt 0 \\
\hline
& & & & & \kern3pt 0 & \kern6pt 0 \kern3pt & \kern3pt 0 & \kern6pt 0 & \kern3pt 1 \\
& & & & & & \kern6pt 1 \kern3pt & \kern3pt 0 & \kern6pt 0 & \kern3pt 0 & 0 \\
& & & & & & & \kern3pt 0 & \kern6pt 0 & \kern3pt 0 & 0 & 1 \\
& & & & & & & & \kern6pt 1 & \kern3pt 0 & 0 & 0 & 0 \\
& & & & & & & & & \ddots & \ddots & \ddots & \ddots & \ddots
\end{array}
\]

Clearly the model is more complicated than it needs to be: we can separate
it into two non-interacting evolutions corresponding to spins up and down.
Consider for instance the set of spins pointing up which move in one unit
of time one step to the right. The corresponding matrix $U_\uparrow$ is
now the doubly infinite matrix
\[
\begin{array}{cccc|ccccc}
\ddots & \ddots & & & & & & & \\
& 0 & 1 & & & & & & \\
& & 0 & \kern3pt 1 \kern2pt \\
& & & \kern3pt 0 \kern2pt & \kern2pt 1 \\
\hline
& & & & \kern2pt 0 & \kern3pt 1 \\
& & & & & \kern3pt 0 & \kern3pt 1 \\
& & & & & & \kern3pt 0 & 1 \\
& & & & & & & \ddots & \ddots
\end{array}
\]

The doubly infinite column vector
$v(z)=(\dots,z^{-2},z^{-1},1,z,z^2,\dots)^T$ clearly satisfies
\[
U_\uparrow v(z) = z v(z).
\]
Since the components $v_j(z)=z^j$, $j\in\mathbb{Z}$, of $v(z)$ are
orthonormal with respect to the Lebesgue measure, arguments entirely
similar to those of the last paragraph of the previous example show that
\[
\int_\mathbb{T} \overline{v_k(z)}\,z^nv_j(z) \frac{dz}{2\pi iz} = (U_\uparrow^n)_{j,k},
\]
an identity which can be trivially checked by a direct computation.

There is another approach to the problem that reduce this example to the
previous one. Since all spins are pointing up we can ignore their
orientation and concentrate on their locations $\dots,-2,-1,0,1,2,\dots$
as a way of describing the pure states. There is now a useful trick (used
in a previous example in Section \ref{CRW}) that consists in ``folding''
the set of all integers in the way
\[
0,-1,1,-2,2,-3,3,\dots
\]
or equivalently relabelling them as follows
\[
\dots, 5, 3, 1, 0, 2 , 4, 6, \dots
\]
With this relabelling the doubly infinite matrix $U_\uparrow$ becomes the
semi-infinite matrix $U_+$ and we are back in the situation discussed in
the previous example.

The evolution of the states with spins pointing down is also described by
the matrix $U_+$ of the previous example, but the required ``folding" in
this case is
\[
-1,0,-2,1,-3,2,-4,\dots
\]

Notice that the initial transition matrix $U$ is the doubly infinite CMV
matrix with null Verblunsky parameters. We will see that common QRWs on the
integers, such as the Hadamard one, essentially correspond to other less
trivial doubly infinite CMV matrices.

The elementary character of the previous examples is the reason to
introduce them early on, since they show in an extremely simple way the
typical features of the method that we will employ in more intrincate
cases. For instance, the decoupling of a QRW on the integers into two QRWs
on the non-negative integers will take place in the next examples too, a
fact that simplifies considerably their analysis. Nevertheless, the
decoupling in the following examples is not so easy to notice since it
holds in a basis of mixed states instead of pure ones.

\section{A more general case: QRWs with distinct coins} \label{GEN}

Here and in the rest of the paper ``distinct coins" means not necessarily
``constant coins".

The previous example on the integers can be generalized to more
interesting QRWs with no decoupling between up and down states, by
including possible transitions between such states. A simple way to do
this is to consider the following dynamics: a spin up can move to the
right and remain up or (finally we get away from deterministic models)
move to the left and change orientation. A spin down can either go to the
right and change orientation or go to the left and remain down.

In other words, only the nearest neighbour transitions such that the final
spin (up/down) agrees with the direction of motion (right/left) are
allowed. This dynamics bears a resemblance to the effect of a magnetic
interaction on quantum system with spin: the spin decides the direction of
motion.

Schematically, the allowed transitions are
\[
\begin{aligned}
|i\>\otimes\uk \longrightarrow \begin{cases}
|i+1\>\otimes\uk & \text{with probability amplitude} \hskip 5pt c_{11}^i
\\
|i-1\>\otimes\dk & \text{with probability amplitude} \hskip 5pt c_{21}^i
\end{cases}
\\
|i\>\otimes\dk \longrightarrow \begin{cases}
|i+1\>\otimes\uk & \text{with probability amplitude} \hskip 5pt c_{12}^i
\\
|i-1\>\otimes\dk & \text{with probability amplitude} \hskip 5pt c_{22}^i
\end{cases}
\end{aligned}
\]
where, for each $i\in\mathbb{Z}$,
\begin{equation} \label{COIN}
C_i = \begin{pmatrix} c_{11}^i & c_{12}^i \\ c_{21}^i & c_{22}^i \end{pmatrix}
\end{equation}
is an arbitrary unitary matrix which we will call the $i^{th}$ coin.

Notice that this is already more general than the Hadamard example usually
discussed in the literature. We will analyze the general case of a
constant quantum coin in more detail in the next section, and then the
popular Hadamard special case in a later section.

The transition matrix $U$ is the unitary doubly infinite matrix
\[
\begin{array}{ccccccc|ccccccc}
\ddots & \ddots & \ddots & \ddots & \ddots & & & \\
& 0 & c^{-2}_{21} & 0 & 0 & c^{-2}_{11} & \\
& & c^{-2}_{22} & 0 & 0 & c^{-2}_{12} & 0 \\
& & & 0 & c^{-1}_{21} & 0 & 0 & c^{-1}_{11} \\
& & & & c^{-1}_{22} & 0 & 0 & c^{-1}_{12} & 0 \\
\hline
& & & & & 0 & c^0_{21} & 0 & 0 & c^0_{11} \\
& & & & & & c^0_{22} & 0 & 0 & c^0_{12} & 0 \\
& & & & & & & 0 & c^1_{21} & 0 & 0 & c^1_{11} \\
& & & & & & & & c^1_{22} & 0 & 0 & c^1_{12} & 0 \\
& & & & & & & & & \ddots & \ddots & \ddots & \ddots & \ddots
\end{array}
\]
which has the structure of a doubly infinite CMV matrix with null odd
Verblunsky parameters
\[
\begin{array}{ccccccc|ccccccc}
\ddots & \ddots & \ddots & \ddots & \ddots & & & \\
& 0 & \overline{\alpha}_{-4} & \kern-5pt 0 & \kern-3pt 0 & \kern-7pt \rho_{-4} & \\
& & \rho_{-4} & \kern-5pt 0 & \kern-3pt 0 & \kern-7pt -\alpha_{-4} & \kern-5pt 0 \\
& & & \kern-5pt 0 & \kern-3pt \overline{\alpha}_{-2} & \kern-7pt 0 & \kern-5pt 0 & \kern-5pt \rho_{-2} \\
& & & & \kern-3pt \rho_{-2} & \kern-7pt 0 & \kern-5pt 0 & \kern-5pt -\alpha_{-2} & \kern-5pt 0 \\
\hline
& & & & & \kern-7pt 0 & \kern-5pt \overline{\alpha}_0 & \kern-5pt 0 & \kern-5pt 0 & \kern-3pt \rho_0 \\
& & & & & & \kern-5pt \rho_0 & \kern-5pt 0 & \kern-5pt 0 & \kern-3pt -\alpha_0 & \kern-3pt 0 \\
& & & & & & & \kern-5pt 0 & \kern-5pt \overline{\alpha}_2 & \kern-3pt 0 & \kern-3pt 0 & \kern-5pt \rho_2 \\
& & & & & & & & \kern-5pt \rho_2 & \kern-3pt 0 & \kern-3pt 0 & \kern-5pt -\alpha_2 & \kern-3pt 0 \\
& & & & & & & & & \ddots & \ddots & \ddots & \ddots & \ddots
\end{array}
\]
However, $U$ is not exactly a CMV matrix unless the diagonal elements of
every coin $C_i$ are positive so that they can be identified with
$\rho_{2i}$.

Notice that $|c_{11}^i|=|c_{22}^i|$ and $|c_{12}^i|=|c_{21}^i|$ due to the
unitarity of $C_i$. The coin $C_i$ will be called trivial when
$c_{11}^i=c_{22}^i=0$ and non trivial otherwise. If some coin is trivial
then the transition matrix $U$ becomes a direct sum of two unitary
matrices. In what follows we will assume that every coin is non trivial,
otherwise the QRW splits directly into independent simpler ones in the
basis of pure states.

Now, a simple change of phases in the basis transforms $U$ in a CMV
matrix: if $e^{i\sigma^j_k}$ is the phase of $c^j_{kk}$ then it is easy to
check that $\mathcal{C}=\Lambda^\dag U \Lambda$ is a doubly infinite CMV
matrix where
$\Lambda=\text{diag}(\dots,\lambda_{-1},\lambda_0,\lambda_1,\dots)$ is
given by $\lambda_{2j+2}=e^{-i\sigma^j_1}\lambda_{2j}$ and
$\lambda_{2j+1}=e^{i\sigma^j_2}\lambda_{2j-1}$ with
$\lambda_{-1}=\lambda_0=1$. The corresponding Verblunsky parameters are
\[
\alpha_{2j} = \overline{c}^j_{21} \frac{\lambda_{2j}}{\lambda_{2j-1}},
\qquad \alpha_{2j+1}=0,
\]
so that $\rho_{2j}=|c_{11}^j|$ are positive.

As in the previous example on all the integers, we can do a folding trick
to transform the doubly infinite transition matrix in a semi-infinite one.
The appropriate ordering of the pure states to obtain a banded
semi-infinite matrix as narrow as possible is
\begin{equation} \label{FOLD-Matr}
\begin{aligned}
& |0\>\otimes\uk,|-1\>\otimes\dk,|-1\>\otimes\uk,|0\>\otimes\dk,
\\
& |1\>\otimes\uk,|-2\>\otimes\dk,|-2\>\otimes\uk,|1\>\otimes\dk,
\\
& \kern12pt \dots \kern42pt \dots \kern42pt \dots \kern42pt \dots
\end{aligned}
\end{equation}
which combines in a suitable way the orderings
\[
|0\>\otimes\uk,|-1\>\otimes\uk, |1\>\otimes\uk,|-2\>\otimes\uk,\dots
\]
and
\[
|-1\>\otimes\dk,|0\>\otimes\dk,|-2\>\otimes\dk,|1\>\otimes\dk,\dots
\]
used in the example discussed in Section \ref{FREE2}. The reason for this
choice is that, in contrast to that case, up and down states do not
decouple now and we must interlace their orderings.

Let us denote by $\bs{U},\bs{\mathcal{C}},\bs{\Lambda}$ the result of
performing such a reordering on $U,\mathcal{C},\Lambda$. Then,
$\bs{\mathcal{C}}=\bs{\Lambda}^\dag\bs{U}\bs{\Lambda}$, with
$\bs{\Lambda}$ diagonal unitary and $\bs{U},\bs{\mathcal{C}}$ with a
$2\times2$-block CMV structure. Moreover, $\bs{\mathcal{C}}$ is exactly
such a block CMV matrix, and its matrix Verblunsky parameters
$(\bs{\alpha}_j)_{j=0}^\infty$ are
\[
\bs{\alpha}_{2j}=\begin{pmatrix} 0 & -\overline{\alpha}_{-2j-2} \\ \alpha_{2j} & 0 \end{pmatrix},
\qquad
\bs{\alpha}_{2j+1}={\bs 0},
\]
where $(\alpha_j)_{j\in\mathbb{Z}}$ are the Verblunsky parameters of
$\mathcal{C}$. Explicitly,
\begin{equation} \label{C2-Matr}
\bs{\mathcal{C}} =
\begin{pmatrix}
\bs{\alpha}_0^\dag & \bs{0} & \bs{\rho}_0^L \\
\bs{\rho}_0^R & \bs{0} & -\bs{\alpha}_0 & \bs{0} \\
\bs{0} & \bs{\alpha}_2^\dag & \bs{0} & \bs{0} & \bs{\rho}_2^L \\
& \bs{\rho}_2^R & \bs{0} & \bs{0} & -\bs{\alpha}_2 & \bs{0} \\
& & \bs{0} & \bs{\alpha}_4^\dag & \bs{0} & \bs{0} & \bs{\rho}_4^L \\
& & & \bs{\rho}_4^R & \bs{0} & \bs{0} & -\bs{\alpha}_4 & \bs{0} \\
& & & & \ddots & \ddots & \ddots & \ddots & \ddots
\end{pmatrix}.
\end{equation}
The positive definite matrices $\bs{\rho}_{2j}^R,\bs{\rho}_{2j}^L$ are
given by
\[
\bs{\rho}_{2j}^R = \begin{pmatrix} \rho_{2j-2} & 0 \\ 0 & \rho_{2j} \end{pmatrix},
\qquad
\bs{\rho}_{2j}^L = \begin{pmatrix} \rho_{2j} & 0 \\ 0 & \rho_{2j-2} \end{pmatrix}
\]

Summarizing, the class of QRWs on the integers with arbitrary non trivial
quantum coins can be described either by using doubly infinite CMV
matrices or $2\times2$-block CMV matrices, in both cases the odd
Verblunsky parameters vanish. The interest of these results is that they
allow us to obtain for some examples of these QRWs a KMcG formula in terms
of matrix valued Szeg\H o polynomials.

Such a KMcG formula comes from the fact that the block CMV matrix
$\bs{\mathcal{C}}$ has an associated vector
$\bs{x}=(\bs{1},\bs{x}_1,\bs{x}_2,\dots)^T$ of left orthonormal Laurent
polynomials satisfying $\bs{\mathcal{C}}\bs{x}(z)=z\bs{x}(z)$. Let
$\Lambda_j$ be the $2\times2$ diagonal blocks of
$\bs{\Lambda}=\text{diag}(\Lambda_0,\Lambda_1,\dots)$. Then,
$\Lambda_0=\bs{1}$ and $\bs{U}\bs{X}(z)=z\bs{X}(z)$ for
$\bs{X}=(\bs{1},\bs{X}_1,\bs{X}_2,\dots)^T$ defined by
$\bs{X}_j=\Lambda_j\bs{x}_j$. Besides, $(\bs{X}_j)_{j=0}^\infty$ is left
orthonormal with respect to the orthogonality measure $\bs{\mu}$ of
$(\bs{x}_j)_{j=0}^\infty$ (i.e., $(\bs{X}_j)_{j=0}^\infty$ and
$(\bs{x}_j)_{j=0}^\infty$ are left orthonormal Laurent polynomials for the
same measure with different normalizations). In consequence,
\[
\begin{aligned}
\int_\mathbb{T} z^n \bs{X}_j(z) d\bs{\mu}(z) \bs{X}_k(z)^\dag
= \sum_{l=0}^\infty \int_\mathbb{T} (\bs{U}^n)_{j,l} \bs{X}_l(z) d\bs{\mu}(z) \bs{X}_k(z)^\dag
= (\bs{U}^n)_{j,k},
\end{aligned}
\]
which can be interpreted as a KMcG formula for the related QRW on the
integers. Here $(\bs{U}^n)_{j,k}$ stands for the $2\times2$ blocks making
up $\bs{U}^n=((\bs{U}^n)_{j,k})_{j,k=0}^\infty$, i.e.,
\begin{equation} \label{PA-Matr}
\begin{aligned}
& (\bs{U}^n)_{2j,2k} =
\begin{pmatrix}
u^n_{j\uparrow,k\uparrow} & u^n_{j\uparrow,-(k+1)\downarrow}
\\
u^n_{-(j+1)\downarrow,k\uparrow} & u^n_{-(j+1)\downarrow,-(k+1)\downarrow}
\end{pmatrix},
\\
& (\bs{U}^n)_{2j,2k+1} =
\begin{pmatrix}
u^n_{j\uparrow,-(k+1)\uparrow} & u^n_{j\uparrow,k\downarrow}
\\
u^n_{-(j+1)\downarrow,-(k+1)\uparrow} & u^n_{-(j+1)\downarrow,k\downarrow}
\end{pmatrix},
\\
& (\bs{U}^n)_{2j+1,2k} =
\begin{pmatrix}
u^n_{-(j+1)\uparrow,k\uparrow} & u^n_{-(j+1)\uparrow,-(k+1)\downarrow}
\\
u^n_{j\downarrow,k\uparrow} & u^n_{j\downarrow,-(k+1)\downarrow}
\end{pmatrix},
\\
& (\bs{U}^n)_{2j+1,2k+1} =
\begin{pmatrix}
u^n_{-(j+1)\uparrow,-(k+1)\uparrow} & u^n_{-(j+1)\uparrow,k\downarrow}
\\
u^n_{j\downarrow,-(k+1)\uparrow} & u^n_{j\downarrow,k\downarrow}
\end{pmatrix},
\end{aligned}
\end{equation}
$u^n_{j,k}$ being the probability amplitude to go from the pure state
$j$ to the pure state $k$ in $n$ steps.

The measure $\bs{\mu}$ and the orthonormal Laurent polynomials
$(\bs{X}_j)_{j=0}^\infty$ in the above KMcG formula will be called the
measure and orthonormal Laurent polynomials associated with the related
QRW on the integers.

\section{QRWs with a constant coin} \label{CON}

The Hadamard QRW is an example of the QRWs described in the previous
section. It corresponds to a constant coin $C_i=H$ given by
\begin{equation} \label{H}
H = \frac{1}{\sqrt{2}} \begin{pmatrix} 1 & 1 \\ 1 & -1 \end{pmatrix}.
\end{equation}
The Hadamard QRW is an example of an unbiased QRW, i.e., a QRW with a
constant coin such that all the allowed transitions are equiprobable.

We will consider in this section a more general class of QRWs on the
integers: those which have an arbitrary constant (unitary) coin
\begin{equation} \label{COIN-CON}
C_i = C = \begin{pmatrix} c_{11} & c_{12} \\ c_{21} & c_{22}
\end{pmatrix}.
\end{equation}
Identifying $C$ with the operator on the spin state space given by
\begin{equation} \label{C-OP}
C\uk = c_{11}\uk + c_{21}\dk, \qquad C\dk = c_{12}\uk + c_{22}\dk,
\end{equation}
the QRW evolution operator can be written $S \otimes C$ with $S$ as in
(\ref{S}).

Let $e^{i\sigma_k}$ be the phase of $c_{kk}$. According to the previous
discussion, the doubly infinite transition matrix is
$U=\Lambda\mathcal{C}\Lambda^\dag$, where
\[
\Lambda=\text{diag}(\dots,\lambda_{-1},\lambda_0,\lambda_1,\dots),
\qquad
\lambda_{2j}=e^{-ij\sigma_1},
\qquad
\lambda_{2j-1}=e^{ij\sigma_2},
\]
and $\mathcal{C}$ is the doubly infinite CMV matrix with Verblunsky
parameters
\[
\alpha_{2j-1}=0, \qquad \alpha_{2j} = \overline{c}_{21} e^{-i2j\vartheta},
\qquad
\vartheta=\frac{1}{2}(\sigma_1+\sigma_2).
\]

The semi-infinite form $\bs{U}$ of this transition matrix can be
factorized as $\bs{U}=\bs{\Lambda}\bs{\mathcal{C}}\bs{\Lambda}^\dag$ with
$\bs{\Lambda}=\text{diag}(\bs{1},\Lambda_1,\Lambda_2,\dots)$ given by
\[
\Lambda_{2j-1} = \begin{pmatrix} e^{ij\sigma_1} & 0 \\ 0 & e^{ij\sigma_2} \end{pmatrix},
\qquad
\Lambda_{2j} = \Lambda_{2j-1}^\dag,
\]
and $\bs{\mathcal{C}}$ the $2\times2$-block CMV matrix with Verblunsky
parameters $\bs{\alpha}_j = A_j e^{-i(j+1)\vartheta}$, where
\[
A_j = \begin{cases} \bs{0} & \text{odd } j, \\ A & \text{even } j, \end{cases}
\qquad
A = \begin{pmatrix} 0 & -\overline{a} \\ a & 0 \end{pmatrix},
\qquad
a = \overline{c}_{21} e^{i\vartheta}.
\]

The Verblunsky parameters $\bs{\alpha}_j$ are simultaneously unitarily
diagonalizable because
\[
P^\dag A P = \begin{pmatrix} i|a| & 0 \\ 0 & -i|a| \end{pmatrix},
\qquad
P = \frac{1}{\sqrt{2}} \begin{pmatrix} 1 & -i\frac{\overline{a}}{|a|} \\ -i\frac{a}{|a|} & 1 \end{pmatrix}.
\]
Hence, we can use the infinite matrix $\bs{P}=\text{diag}(P,P,P,\dots)$ to
transform $\bs{\mathcal{C}}$ into a new block CMV matrix
$\bs{\mathcal{C}}^{(0)} = \bs{P}^\dag \bs{\mathcal{C}} \bs{P}$ with
diagonal Verblunsky parameters $\bs{\alpha}_j^{(0)}$ given by
\[
\bs{\alpha}_j^{(0)} = \begin{pmatrix} \alpha_j^{(0)} & 0 \\ 0 & -\alpha_j^{(0)} \end{pmatrix},
\qquad
\alpha_j^{(0)}=a_je^{-i(j+1)\vartheta},
\]
where
\begin{equation} \label{VP-DIAG}
a_j = \begin{cases} 0 & \text{odd } j, \\ i|a| & \text{even } j. \end{cases}
\end{equation}

This means that, although the pure states do not decouple, there is a
basis in which the QRW decouples in two independent ones corresponding to
scalar CMV matrices with Verblunsky parameters $\pm\alpha_j^{(0)}$.

More precisely, let $x^\pm_j$ and $\mu_\pm$ be the orthonormal Laurent
polynomials and measure associated with the Verblunsky parameters
$\pm\alpha_j^{(0)}$. Since
$\bs{U}=\bs{Q}\bs{\mathcal{C}}^{(0)}\bs{Q}^\dag$, $\bs{Q}=\bs{\Lambda P}$,
we find that $\bs{U}\bs{X}(z)=z\bs{X}(z)$ for
$\bs{X}=(\bs{1},\bs{X}_1,\bs{X}_2,\dots)^T$ defined by
\[
\bs{X}_j = \Lambda_j \bs{x}_j,
\qquad
\bs{x}_j = P \begin{pmatrix} x^+_j & 0 \\ 0 & x^-_j \end{pmatrix} P^\dag,
\]
and $\bs{X}_j$ are left orthonormal with respect to the same matrix
measure $\bs{\mu}$ as $\bs{x}_j$, that is,
\[
\bs{\mu} = P \begin{pmatrix} \mu_+ & 0 \\ 0 & \mu_- \end{pmatrix} P^\dag.
\]
Therefore,
\begin{equation} \label{KMG-CON-LINE}
\int_\mathbb{T} z^n\bs{X}_j(z)\,d\bs{\mu}(z)\,\bs{X}_k(z)^\dag = (\bs{U}^n)_{j,k},
\end{equation}
which will yield a KMcG formula for the QRW with a constant coin on the
integers once the measure $\bs{\mu}$ and the orthonormal Laurent
polynomials $\bs{X}_j$ are obtained in Section \ref{CON-LINE}.

So far, we have split a QRW on the integers with a constant coin into two
QRWs on the non-negative integers. Now, to obtain a KMcG formula we just
need to compute the scalar orthonormal Laurent polynomials $x^\pm_j$ and
the corresponding scalar orthogonality measures $\mu_\pm$. Moreover, the
corresponding Verblunsky coefficients are $\pm a_je^{-i(j+1)\vartheta}$
with $a_j$ given in (\ref{VP-DIAG}), hence (\ref{ROT-VP}) shows that
$\mu_\pm$ can be obtained rotating by an angle $\vartheta$ the measure
with Verblunsky parameters
\begin{equation} \label{i|a|0}
\qquad\qquad\quad\pm i|a|, \; 0, \; \pm i|a|, \; 0, \; \pm i|a|, \; 0, \; \dots
\qquad\quad (|a|=|c_{21}|)
\end{equation}

Explicitly, from (\ref{ROT-VP}) and (\ref{ROT-OLP}) we find that
\begin{equation} \label{ME}
\begin{aligned}
& d\bs{\mu}(z) = d\hat{\bs{\mu}}(e^{-i\vartheta}z),
\qquad
\hat{\bs{\mu}} = P \begin{pmatrix} \hat{\mu}_+ & 0 \\ 0 & \hat{\mu}_- \end{pmatrix} P^\dag,
\\
& \bs{X}_j(z) = \hat{\Lambda}_j \hat{\bs{x}}_j(e^{-i\vartheta}z),
\qquad
\hat{\bs{x}}_j = P \begin{pmatrix} \hat{x}^+_j & 0 \\ 0 & \hat{x}^-_j \end{pmatrix} P^\dag,
\\
& \hat{\Lambda}_{2j-1} = \begin{pmatrix} e^{ij(\sigma_1-\sigma_2)/2} & 0 \\ 0 & e^{ij(\sigma_2-\sigma_1)/2} \end{pmatrix},
\qquad \hat{\Lambda}_{2j}=\hat{\Lambda}_{2j-1}^\dag,
\end{aligned}
\end{equation}
where $\hat{x}^\pm_j$ and $\hat{\mu}_\pm$ are the orthonormal Laurent
polynomials and measures with Verblunsky parameters (\ref{i|a|0}). Hence,
the matrix Carath\'{e}odory function $\bs{F}$ of $\bs{\mu}$ is related to the
Carath\'{e}odory functions $\hat{F}_\pm$ of $\hat{\mu}_\pm$ by
\begin{equation} \label{ME-CF}
\bs{F}(z) = \hat{\bs{F}}(e^{-i\vartheta}z),
\qquad
\hat{\bs{F}} = P \begin{pmatrix} \hat{F}_+ & 0 \\ 0 & \hat{F}_- \end{pmatrix} P^\dag.
\end{equation}

A CMV matrix with the Verblunsky parameters (\ref{i|a|0}) can be
understood as the transition matrix of a QRW with a constant coin on the
non-negative integers. Thus, a QRW with a constant coin on the integers
always splits in two QRWs with constant coins on the non-negative
integers. Such a splitting is not initially obvious since it takes place
in a basis of mixed states.

It is worth to remark that the block CMV matrix related to a constant coin
$C$ on the integers depends only on the modulus of the entries of $C$ (one
of them determines the others by unitarity). This means that QRWs on the
integers with a constant coin do not depend essentially on the phases of
its entries. For an unbiased QRW all the entries of the coin $C$ have
equal modulus, so $|a|=|c_{21}|=1/\sqrt{2}$ due to the unitarity of $C$.
The Hadamard QRW is therefore just a canonical example of unbiased QRW on
the integers.

Bearing in mind that we are reducing the problem on the integers to two
problems on the non-negative integers, in the next section we will discuss
completely the QRWs with a constant coin on the non-negative integers.

\section{QRWs with a constant coin on the non-negative integers} \label{CON-HALF}

QRWs with a constant coin on the non-negative integers are a preliminary
step to complete the discussion of QRWs on the integers. Besides, these
QRWs have their own interest because they present special features that do
not appear for QRWs on the integers.

Let us suppose that we order the pure states on the non-negative integers
as in (\ref{ORD}). A unitary matrix like
\[
U = \begin{pmatrix}
c^0_{21} & 0 & c^0_{11} \\
c^0_{22} & 0 & c^0_{12} & 0 \\
0 & c^1_{21} & 0 & 0 & c^1_{11} \\
& c^1_{22} & 0 & 0 & c^1_{12} & 0 \\
& & 0 & c^2_{21} & 0 & 0 & c^2_{11} \\
& & & c^2_{22} & 0 & 0 & c^2_{12} & 0 \\
& & & & \ddots & \ddots & \ddots & \ddots & \ddots
\end{pmatrix}
\]
can be understood as the transition matrix for a QRW on the non-negative
integers with arbitrary (unitary) coins $C_i$ as in (\ref{COIN}) for
$i=0,1,2,\dots$.

A constant coin $C$ corresponds to the operator $S_+ \otimes C$ where
$S_+$ is given in (\ref{S+}) and $C$ is identified with the operator
(\ref{C-OP}).

Exactly as in to Section \ref{GEN}, even in the case of (non trivial)
distinct coins, a simple change of phases in the basis reduces $U$ to a
scalar CMV matrix with null odd Verblunsky coefficients:
$\mathcal{C}=\Lambda^\dag U \Lambda$ is CMV for
$\Lambda=\text{diag}(\lambda_0,\lambda_1,\dots)$ given by
$\lambda_{2j+2}=e^{-i\sigma^j_1}\lambda_{2j}$,
$\lambda_{2j+1}=e^{i\sigma^j_2}\lambda_{2j-1}$ with
$\lambda_{-1}=\lambda_0=1$ and $e^{i\sigma^j_k}$ the phase of $c^j_{kk}$.
Its Verblunsky parameters are given again by
\[
\alpha_{2j} = \overline{c}^j_{21} \frac{\lambda_{2j}}{\lambda_{2j-1}},
\qquad \alpha_{2j+1}=0,
\]
so that $\rho_{2j}=|c_{11}^j|$ are positive.

If $\mu$ and $x_j$ are the measure and orthonormal Laurent polynomials
associated with $\mathcal{C}$, then $UX(z)=zX(z)$ with
$X=(1,X_1,X_2,\dots)^T$ and $X_j=\lambda_jx_j$. Since $\mu$ makes the
polynomials $X_j$ orthonormal too,
\begin{equation} \label{KMG-CON-HALF}
\int_\mathbb{T} z^nX_j(z)\,\overline{X_k(z)}\,d\mu(z) = (U^n)_{j,k},
\end{equation}
which is a KMcG formula for a QRW with arbitrary distinct coins on the
non-negative integers.

The above formula simply says that the transition matrix $U$ has been
identified as a matrix representation of the unitary multiplication
operator $U_\mu$ on $L^2_\mu(\mathbb{T})$ given by (\ref{UMO}).

The measure $\mu$ and the orthonormal Laurent polynomials $X_j$ in
(\ref{KMG-CON-HALF}) will be called the measure and orthonormal Laurent
polynomials associated with the related QRW on the non-negative integers.

Following the arguments and notation of Section \ref{CON} we can see that
a constant coin $C$ leads to a rotation of the case associated with a CMV
matrix with Verblunsky parameters
\begin{equation} \label{a0}
\qquad\qquad\qquad\qquad a, \; 0, \; a, \; 0, \; a, \; 0, \; \dots
\qquad\qquad\quad (a=\overline{c}_{21}e^{i\vartheta})
\end{equation}
In contrast to (\ref{i|a|0}), the even Verblunsky parameters in (\ref{a0})
can have an arbitrary phase which comes from the phases of the
coefficients of the coin. Hence, we should expect a stronger influence of
the phases of the coin for constant coins on the non-negative integers
compared to the same situation on the integers.
This will
bring in new possibilities when discussing the
semi-infinite version of the Hadamard QRW, corresponding to
$|a|=1/\sqrt{2}$.

To be more precise, if $\Lambda=\text{diag}(\lambda_0,\lambda_1,\dots)$,
$\lambda_{2j}=e^{-ij\sigma_1}$, $\lambda_{2j-1}=e^{ij\sigma_2}$, then
$\mathcal{C}=\Lambda^\dag U \Lambda$ is a CMV matrix whose measure $\mu$
is obtained by rotating by an angle $\vartheta$ the measure $\hat{\mu}$
with Verblunsky parameters (\ref{a0}). If $x_j$ and $\hat x_j$ are the
orthonormal Laurent polynomials associated with $\mu$ and $\hat\mu$
respectively, we know from (\ref{ROT-OLP}) that
$x_{2j-1}(z)=e^{-ij\vartheta}\hat x_{2j-1}(e^{-i\vartheta}z)$ and
$x_{2j}(z)=e^{ij\vartheta}\hat x_{2j}(e^{-i\vartheta}z)$. So, according
to the previous discussion, the Laurent polynomials
\begin{equation}
X_j(z) = \lambda_jx_j(z) = \hat{\lambda}_j\,\hat{x}_j(e^{-i\vartheta}z),
\qquad \hat{\lambda}_{2j-1} = \hat{\lambda}_{2j} =
e^{ij(\sigma_2-\sigma_1)/2},
\end{equation}
are orthonormal with respect to $\mu$, satisfy $UX(z)=zX(z)$,
$X=(1,X_1,X_2,\dots)^T$, and provide the KMcG formula
(\ref{KMG-CON-HALF}), which can be rewritten as
\begin{equation} \label{KMG-CON-HALF-2}
(U^n)_{j,k} = e^{in\vartheta} \frac{\hat{\lambda}_j}{\hat{\lambda}_k}
\int_\mathbb{T} z^n\hat{x}_j(z)\,\overline{\hat{x}_k(z)}\,d\hat{\mu}(z).
\end{equation}

To get from (\ref{KMG-CON-HALF}) a KMcG formula for the QRWs with a
constant coin on the non-negative integers we simply need $\hat{x}_j$ and
$\hat{\mu}$, which are calculated in an appendix at the end of the paper.
The main results from the appendix are summarized now:

 An expression for the orthonormal Laurent polynomials is given by
\[
\begin{aligned}
\hat{x}_{2j-1}(z) = U_j(y)-\rho^{-1}(z+a)U_{j-1}(y),
\\
\hat{x}_{2j}(z) = U_j(y)-\rho^{-1}(z^{-1}+\overline{a})U_{j-1}(y),
\end{aligned}
\qquad y=\frac{1}{2\rho}(z+z^{-1}),
\]
with $\rho=\sqrt{1-|a|^2}$ and $U_j$ the second kind Chebyshev polynomials
given by (\ref{eq2}). In particular,
$\hat{x}_{2j}(z)=\overline{\hat{x}_{2j-1}(1/\overline{z})}$.

On the other hand, the measure is given by
\[
\begin{aligned}
& d\hat{\mu}(e^{i\theta}) =
w(\theta)\frac{d\theta}{2\pi} +
M\delta(\theta-\beta)d\theta,
\\
& w(\theta) =
\frac{\sqrt{\sin^2\theta-\sin^2\eta}}{|\sin\theta-\sin\beta|},
\qquad
\theta\in[\eta,\pi-\eta]\cup[\eta-\pi,-\eta],
\\
& M=\frac{|\re a|}{\sqrt{1-|\im a|^2}} =
\frac{\sqrt{\sin^2\eta-\sin^2\beta}}{|\cos\beta|},
\end{aligned}
\]
where the angles $\eta\in[0,\pi/2)$ and $\beta\in(-\pi,\pi]$ are defined
by means of
\[
\sin\eta=|a|,
\qquad
\sin\beta=-\im a,
\qquad
\text{sign}(\cos\beta)=\text{sign}(\re a).
\]

The corresponding Carath\'{e}odory function is
$F(z)=\hat{F}(e^{-i\vartheta}z)$ where
\[
\hat{F}(z) \kern-2pt = \kern-2pt
-\frac{\sqrt{(z-z^{-1})^2+4|a|^2}+2\re a}{z-z^{-1}+2i\im a}
\kern-2pt = \kern-2pt
-\frac{z-z^{-1}-2i\im a}{\sqrt{(z-z^{-1})^2+4|a|^2}-2\re a}.
\]

Now that we have all the ingredients that enter in the integral in
(\ref{KMG-CON-HALF-2}), notice that the only parameter that appears in it
is the value of the complex number $a$. The parameter $\vartheta$ as well
as the individual values of $\sigma_1,\sigma_2$ have an effect on
$(U^n)_{j,k}$ but this appears only as a factor in front of the integral
on the right hand side.

The measure $d\mu(z)=d\hat{\mu}(e^{-i\vartheta}z)$ appearing in
(\ref{KMG-CON-HALF}) has in general a continuous weight plus a Dirac
delta. The weight is supported on two symmetric arcs of angular amplitude
$2\eta$ centered at $\pm ie^{i\vartheta}$. The mass point is located
outside of the support of the weight and it is absent only when $a$ is
imaginary, which holds exactly when the coin is symmetric
($c_{12}=c_{21}$).

While the location of the weight only depends on $|a|=|c_{21}|$, its form
and the location and mass of the Dirac delta also depend on the phase of
$a=\overline{c}_{21}e^{i\vartheta}$. Therefore, the phases of the coin $C$
have a remarkable influence in the semi-infinite QRWs with a constant
coin, in marked contrast to the case of the integers.

\section{QRWs with a constant coin on the integers} \label{CON-LINE}

The results of the previous section permit us to complete the analysis of
the QRWs with a constant coin on the integers. Let us remember the
notation: $C=(c_{jk})_{j,k=1,2}$ is the constant coin, $e^{i\sigma_k}$ the
phase of $c_{kk}$, $\vartheta=(\sigma_1+\sigma_2)/2$ the angle of
rotation, $a=\overline{c}_{21}e^{i\vartheta}$ and $\rho=\sqrt{1-|a|^2}$.

The matrix ingredients $\bs{X}_j$, $\bs{\mu}$ of the KMcG formula
(\ref{KMG-CON-LINE}) are given by (\ref{ME}) in terms of the scalar ones
$\hat{x}^\pm_j$, $\hat{\mu}_\pm$ with Verblunsky parameters $\pm i|a|, \,
0, \, \pm i|a|, \, 0, \, \dots$, a specialization for $a$ imaginary of the
case $a, \, 0, \, a, \, 0, \, \dots$ analyzed in the appendix. This is
precisely the case where the mass point disappears, so we can anticipate
that the matrix measure for a constant coin on the integers is always
given exclusively by a continuous matrix weight.

Combining the results of Section \ref{CON} and the appendix we get the
following expressions for the scalar objects
\[
\begin{array}{l}
\ds \hat{F}_\pm(z)
=-\frac{\sqrt{(z-z^{-1})^2+4|a|^2}}{z-z^{-1}\pm2i|a|}
=-\frac{z-z^{-1}\mp2i|a|}{\sqrt{(z-z^{-1})^2+4|a|^2}},
\bigskip \\
\ds d\hat{\mu}_\pm(e^{i\theta}) = w_\pm(\theta)\frac{d\theta}{2\pi}, \qquad
w_\pm(\theta)=\sqrt{\frac{\sin\theta\mp\sin\eta}{\sin\theta\pm\sin\eta}},
\smallskip \\
\theta\in[\eta,\pi-\eta]\cup[\eta-\pi,-\eta],
\qquad
\sin\eta=|a|, \qquad \eta\in[0,\pi/2),
\bigskip \medskip \\
\begin{aligned}
& x^\pm_{2j-1}(z) = U_j(y)-\rho^{-1}(z \pm i|a|)U_{j-1}(y),
\smallskip \\
& x^\pm_{2j}(z) = \overline{x_{2j-1}(1/\overline{z})},
\end{aligned}
\ds \qquad y=\frac{1}{2\rho}(z+z^{-1}),
\end{array}
\]
so that the matrix objects in (\ref{ME}) and (\ref{ME-CF}) are given by
\[
\begin{array}{c}
\bs{F}(z) = \hat{\bs{F}}(e^{-i\vartheta}z),
\medskip \\
\ds \hat{\bs{F}}(z)=\frac{-1}{\sqrt{(z-z^{-1})^2+4|a|^2}}
\begin{pmatrix}
z-z^{-1} & 2\overline{a} \\ -2a & z-z^{-1}
\end{pmatrix},
\end{array}
\]
\[
\begin{array}{c}
\ds d\bs{\mu}(z) = d\hat{\bs{\mu}}(e^{-i\vartheta}z),
\qquad
d\hat{\bs{\mu}}(e^{i\theta}) = W(\theta) \frac{d\theta}{2\pi},
\medskip \\
\ds W(\theta) = \frac{1}{\sqrt{\sin^2\theta-\sin^2\eta}}
\begin{pmatrix}
|\sin\theta| & \mp i \overline{a}
\\
\pm i a & |\sin\theta|
\end{pmatrix}
\quad \text{if } \begin{cases} \theta\in[\eta,\pi-\eta], \\ \theta\in[\eta-\pi,-\eta], \end{cases}
\end{array}
\]
\[
\begin{aligned}
& \bs{X}_j(z)=\hat{\Lambda}_j\hat{\bs{x}}_j(e^{-i\vartheta}z),
\qquad
\hat{\Lambda}_{2j-1} = \begin{pmatrix} e^{ij(\sigma_1-\sigma_2)/2} & 0 \\ 0 & e^{ij(\sigma_2-\sigma_1)/2} \end{pmatrix}
= \hat{\Lambda}_{2j}^\dag,
\smallskip \\
& \hat{\bs{x}}_{2j-1}(z) = \bs{1} U_j(y) - \frac{1}{\rho} \begin{pmatrix} z & -\overline{a} \\ a & z \end{pmatrix} U_{j-1}(y),
\qquad
\hat{\bs{x}}_{2j}(z) = \hat{\bs{x}}_{2j-1}(1/\overline{z})^\dag.
\end{aligned}
\]

The expression for the polynomials $\bs{X}_j$ can be compared to the one
at the end of Section \ref{CRW}.

\section{The Hadamard QRW versus other unbiased QRWs} \label{HAD}

The Hadamard QRW is the unbiased QRW on the integers with constant coin
(\ref{H}). Applying the results of the previous section to this case gives
\[
\begin{aligned}
& \bs{F}(z)=\frac{1}{\sqrt{1+z^4}}
\begin{pmatrix}
1+z^2 & \sqrt{2}z \\ \sqrt{2}z & 1+z^2
\end{pmatrix},
\\
& d\bs{\mu}(e^{i\theta}) = \frac{1}{\sqrt{\cos2\theta}}
\begin{pmatrix}
\sqrt{1+\cos2\theta} & \pm1
\\
\pm1 & \sqrt{1+\cos2\theta}
\end{pmatrix}
\frac{d\theta}{2\pi}
\quad \text{if } \begin{cases} \theta\in[-\frac{\pi}{4},\frac{\pi}{4}], \\ \theta\in[\frac{3\pi}{4},\frac{5\pi}{4}], \end{cases}
\\
& \bs{X}_{2j-1}(z) =  \begin{pmatrix} (-i)^j & 0 \\ 0 & i^j \end{pmatrix}
\left\{
\bs{1} U_j(y^*) + i \begin{pmatrix} \sqrt{2}z & -1 \\ -1 & \sqrt{2}z \end{pmatrix} U_{j-1}(y^*)
\right\},
\\
& \bs{X}_{2j}(z) = \bs{X}_{2j-1}(1/z),
\end{aligned}
\]
where, here and below,
\[
y^*=\frac{z-z^{-1}}{\sqrt{2}i},
\]
and the square root $\sqrt{1+z^4}$ is the analytic branch with value 1 at
the origin. Notice that the measure $\bs{\mu}$ is symmetric with respect
to the real line and the matrix coefficients of the Laurent polynomials
$\bs{X}_j$ are real.

We can compare this case with a semi-infinite version with the same
constant coin (\ref{H}). Taking into account Section \ref{CON-HALF}, the
coin (\ref{H}) yields on the non-negative integers
\[
\begin{aligned}
& F(z) = \frac{\sqrt{1+z^4}}{1-\sqrt{2}z+z^2} = \frac{1+\sqrt{2}z+z^2}{\sqrt{1+z^4}},
\\
& d\mu(e^{i\theta}) = \frac{\sqrt{1+\cos2\theta}\pm1}{\sqrt{\cos2\theta}} \, \frac{d\theta}{2\pi}
\quad \text{if } \begin{cases} \theta\in[-\frac{\pi}{4},\frac{\pi}{4}], \\ \theta\in[\frac{3\pi}{4},\frac{5\pi}{4}], \end{cases}
\\
& X_{2j-1}(z) = i^j \{U_j(y^*)+i(\sqrt{2}z-1)U_{j-1}(y^*)\},
\\
& X_{2j}(z) = (-1)^j X_{2j-1}(1/z).
\end{aligned}
\]
We can observe the similarity between the Hadamard QRW and its version on
the non-negative integers. Again, the measure $\mu$ is given by a
symmetric weight and the Laurent polynomials $X_j$ have real coefficients.

However, this similarity goes away for other unbiased QRWs. Consider for
instance the constant equiprobable coin
\begin{equation} \label{Hmod}
C = \frac{1}{\sqrt{2}} \begin{pmatrix} 1 & -i \\ i & -1 \end{pmatrix}.
\end{equation}
Section \ref{CON-LINE} gives for the corresponding QRW on the integers
\[
\begin{aligned}
& \bs{F}(z)=\frac{1}{\sqrt{1+z^4}}
\begin{pmatrix}
1+z^2 & i\sqrt{2}z \\ -i\sqrt{2}z & 1+z^2
\end{pmatrix},
\\
& d\bs{\mu}(e^{i\theta}) = \frac{1}{\sqrt{\cos2\theta}}
\begin{pmatrix}
\sqrt{1+\cos2\theta} & \pm i
\\
\mp i & \sqrt{1+\cos2\theta}
\end{pmatrix}
\frac{d\theta}{2\pi}
\quad \text{if }
\begin{cases}
\theta\in[-\frac{\pi}{4},\frac{\pi}{4}],
\\
\theta\in[\frac{3\pi}{4},\frac{5\pi}{4}],
\end{cases}
\\
& \bs{X}_{2j-1}(z) =  \begin{pmatrix} (-i)^j & 0 \\ 0 & i^j \end{pmatrix}
\left\{
\bs{1} U_j(y^*) + i \begin{pmatrix} \sqrt{2}z & -i \\ i & \sqrt{2}z \end{pmatrix} U_{j-1}(y^*)
\right\},
\\
& \bs{X}_{2j}(z) = \bs{X}_{2j-1}(1/z).
\end{aligned}
\]
As in the Hadamard QRW, the measure $\bs{\mu}$ is given by a weight which
is symmetric with respect to the real line.

From Section \ref{CON-HALF}, the same constant coin (\ref{Hmod}) yields
for the non-negative integers
\[
\begin{aligned}
& F(z) = \frac{\sqrt{1+z^4}-i\sqrt{2}z}{1+z^2} = \frac{1+z^2}{\sqrt{1+z^4}+i\sqrt{2}z},
\\
& d\mu(e^{i\theta}) = \sqrt{\frac{\cos2\theta}{1+\cos2\theta}}\,\frac{d\theta}{2\pi}
+ \frac{1}{\sqrt{2}} \ts \delta(\theta-\frac{\pi}{2})d\theta,
\quad \theta\in[-\frac{\pi}{4},\frac{\pi}{4}]\cup[\frac{3\pi}{4},\frac{5\pi}{4}],
\\
& X_{2j-1}(z) = i^j \{U_j(y^*)+i(\sqrt{2}z+i)U_{j-1}(y^*)\},
\\
& X_{2j}(z) = (-1)^j \overline{X_{2j-1}(1/\overline{z})}.
\end{aligned}
\]

The differences between the infinite and the semi-infinite versions of
this unbiased QRW are evident. On the non-negative integers the weight is
symmetric with respect to the real line, but there is also a mass point
which destroys such a symmetry for the full measure.

The similarities and differences observed above regarding the ingredients
of the KMcG formulas for different QRWs reveal how close or far their
probabilistic behaviours are from each other. To make this more evident
let us obtain the probability amplitudes $u^n_{j,k}$ for a $n$-step
transition between certain states $j,k$ in the different examples.

For the examples on the integers, according to (\ref{PA-Matr}),
\[
(\bs{U}^n)_{0,0} =
\begin{pmatrix}
u^n_{0\uparrow,0\uparrow} & u^n_{0\uparrow,-1\downarrow}
\\
u^n_{-1\downarrow,0\uparrow} & u^n_{-1\downarrow,-1\downarrow}
\end{pmatrix}.
\]
Due to the translation invariance of any QRW with a constant coin on the
integers, $(\bs{U}^n)_{0,0}$ gives the probability amplitudes
$u^n_{k\uparrow,k\uparrow}$, $u^n_{k\downarrow,k\downarrow}$,
$u^n_{k\uparrow,(k-1)\downarrow}$ and $u^n_{k\downarrow,(k+1)\uparrow}$
for any $k$.

On the other hand, the KMcG formula (\ref{KMG-CON-LINE}) states that
$(\bs{U}^n)_{0,0}=\int_\mathbb{T}z^n\,d\bs{\mu}(z)=\bs{\mu}_n$ are the
moments of the related measure, which, following (\ref{MOM-Matr}), are
provided by the McLaurin series of the Carath\'{e}odory function $\bs{F}$.

For the Hadamard QRW on the integers the expression of $\bs{F}$ gives,
apart from the trivial moment $\bs{\mu}_0=\bs{1}$,
\[
\bs{\mu}_{4m}=\bs{\mu}_{4m+2}=\frac{c_m}{2}\bs{1},
\qquad
\bs{\mu}_{4m+1}=\frac{c_m}{\sqrt{2}}\begin{pmatrix}0&1\\1&0\end{pmatrix},
\qquad
\bs{\mu}_{4m+3}=\bs{0},
\]
where
\begin{equation} \label{cn}
c_0=1,
\qquad
\quad c_n = (-1)^n\prod_{k=1}^n\left(1-\frac{1}{2k}\right), \quad n\ge1,
\end{equation}
are the coefficients of the series
$1/\sqrt{1+z}=\sum_{n=0}^\infty c_nz^n$. Hence,
\[
\begin{array}{l}
u^n_{k\uparrow,k\uparrow}=u^n_{k\downarrow,k\downarrow}=
\begin{cases}
\frac{c_m}{2} & \scriptstyle n=4m,4m+2,
\\
0 & \scriptstyle n=2m+1,
\end{cases}
\medskip \\
u^n_{k\uparrow,(k-1)\downarrow}=u^n_{k\downarrow,(k+1)\uparrow}=
\begin{cases}
\frac{c_m}{\sqrt{2}} & \scriptstyle n=4m+1,
\\
0 & \scriptstyle n=2m,4m+3,
\end{cases}
\end{array}
\]

The non-diagonal elements of
\[
(\bs{U}^n)_{1,0} =
\begin{pmatrix}
u^n_{-1\uparrow,0\uparrow} & u^n_{-1\uparrow,-1\downarrow}
\\
u^n_{0\downarrow,0\uparrow} & u^n_{0\downarrow,-1\downarrow}
\end{pmatrix},
\]
permit us to complete the description of the $n$-step transition
amplitudes between the spin states at the same site. From
(\ref{KMG-CON-LINE}) and the expression of the polynomials $\bs{X}_k$ for
the Hadamard QRW we get
\[
\begin{aligned}
(\bs{U}^n)_{1,0} &= \int_\mathbb{T} \bs{X}_1(z) z^n d\bs{\mu} =
\int_\mathbb{T} \left\{ \begin{pmatrix}0&-1\\1&0\end{pmatrix} + \sqrt{2} \begin{pmatrix}1&0\\0&-1\end{pmatrix} z^{-1} \right\}
z^n d\bs{\mu}
\\
&= \begin{pmatrix}0&-1\\1&0\end{pmatrix} \bs{\mu}_n + \sqrt{2} \begin{pmatrix}1&0\\0&-1\end{pmatrix} \bs{\mu}_{n-1},
\end{aligned}
\]
thus, the values of the Hadamard moments give
\[
\begin{array}{l}
u^n_{k\uparrow,k\downarrow}=-u^n_{k\downarrow,k\uparrow}=
\begin{cases}
-\frac{c_m}{2} & \scriptstyle n=4m,
\\
\frac{c_m}{2} & \scriptstyle n=4m+2,
\\
0 & \scriptstyle n=2m+1,
\end{cases}
\medskip \\
u^n_{k\uparrow,(k+1)\uparrow}=-u^n_{k\downarrow,(k-1)\downarrow}=
\begin{cases}
0 & \scriptstyle n=2m,4m+1,
\\
\frac{c_m}{\sqrt{2}} & \scriptstyle n=4m+3,
\end{cases}
\end{array}
\]
except for $u^1_{k\uparrow,(k+1)\uparrow}=-u^1_{k\downarrow,(k-1)\downarrow}=1/\sqrt{2}$.

In the Hadamard example, all the $n$-step transitions between pure spin
states for the same site are forbidden for odd $n$ (except for $n=1$),
while for even $n$ we find probabilities $|c_m|^2/4$ if $n=4m,4m+2$. The
same result holds for any other unbiased QRW on the integers due to the
similarity between the corresponding Carath\'{e}odory functions
and orthonormal Laurent polynomials.

Consider now the Hadamard coin on the non-negative integers. From the
McLaurin series of the associated Carath\'{e}odory function and (\ref{MOM}) we
find that, apart from $\mu_0=1$, the related moments are
\[
\mu_{4m} = \mu_{4m+2} = \frac{c_m}{2},
\qquad
\mu_{4m+1}=\frac{c_m}{\sqrt{2}},
\qquad
\mu_{4m+3}=0.
\]

The KMcG formula (\ref{KMG-CON-HALF}) and the expression
$X_1(z)=1-\sqrt{2}z^{-1}$ for the first orthonormal Laurent polynomial
yield
\[
\begin{array}{l}
(U^n)_{0,0}=\mu_n,
\qquad
(U^n)_{1,1}=3\mu_n-\sqrt{2}\mu_{n+1}-\sqrt{2}\mu_{n-1},
\medskip \\
(U^n)_{1,0}=\mu_n-\sqrt{2}\mu_{n-1},
\qquad
(U^n)_{0,1}=\mu_n-\sqrt{2}\mu_{n+1}.
\end{array}
\]
Therefore,
\[
\begin{array}{l}
u^n_{0\uparrow,0\uparrow}=
\begin{cases}
\frac{c_m}{2} & \scriptstyle n=4m,4m+2,
\\
\frac{c_m}{\sqrt{2}} & \scriptstyle n=4m+1,
\\
0 & \scriptstyle n=4m+3,
\end{cases}
\qquad\kern4pt
u^n_{0\downarrow,0\downarrow}=
\begin{cases}
\frac{c_m}{2} & \scriptstyle n=4m,4m+2,
\\
\frac{c_m}{\sqrt{2}} & \scriptstyle n=4m+1,
\\
-\frac{c_m+c_{m+1}}{\sqrt{2}} & \scriptstyle n=4m+3,
\end{cases}
\medskip \\
u^n_{0\uparrow,0\downarrow}=
\begin{cases}
-\frac{c_m}{2} & \scriptstyle n=4m,
\\
0 & \scriptstyle n=4m+1,
\\
\frac{c_m}{2} & \scriptstyle n=4m+2,
\\
-\frac{c_{m+1}}{\sqrt{2}} & \scriptstyle n=4m+3,
\end{cases}
\qquad
u^n_{0\downarrow,0\uparrow}=
\begin{cases}
\frac{c_m}{2} & \scriptstyle n=4m,
\\
0 & \scriptstyle n=4m+1,
\\
-\frac{c_m}{2} & \scriptstyle n=4m+2,
\\
-\frac{c_m}{\sqrt{2}} & \scriptstyle n=4m+3,
\end{cases}
\end{array}
\]
except for $u^1_{0\downarrow,0\downarrow}=0$ and
$u^1_{0\downarrow,0\uparrow}=-1/\sqrt{2}$.

The above results show that, concerning the $n$-step transitions between
pure spin states at site 0, the Hadamard coin on the non-negative integers
has the same probability amplitudes as the Hadamard coin on the integers
for even $n$, while for odd $n$ some of the transitions remain
forbidden.

On the other hand, the Carath\'{e}odory function for the unbiased QRW on the
non-negative integers with coin (\ref{Hmod}) yields, apart from $\mu_0=1$,
the moments
\[
\mu_{4m} = \frac{d_m}{2},
\qquad
\mu_{4m+1}=\frac{i}{\sqrt{2}},
\qquad
\mu_{4m+2} = -\frac{d_m}{2},
\qquad
\mu_{4m+3}=-\frac{i}{\sqrt{2}},
\]
where $d_n$ are the coefficients of $\sqrt{1+z}/(1-z)=\sum_{n=0}^\infty
d_n z^n$, so that $\sqrt{1+z^2}/(1+z) = \sum_{n=0}^\infty d_n
(z^{2n}-z^{2n+1})$. Explicitly,
\begin{equation} \label{dn}
\begin{array}{c}
\displaystyle d_n=\hat c_0+\hat c_1+\dots+\hat c_n,
\qquad\quad
\sqrt{1+z} = \sum_{n=0}^\infty \hat c_n z^n,
\medskip \\
\displaystyle \hat c_0=1, \qquad \hat c_1=\frac{1}{2},
\qquad
\hat c_n=\frac{(-1)^{n-1}}{2}\prod_{k=2}^n\left(1-\frac{3}{2k}\right), \quad n\ge2.
\end{array}
\end{equation}

Introducing $X_1(z)=-(i+\sqrt{2}z^{-1})$ in the KMcG formula
(\ref{KMG-CON-HALF}) gives
\[
\begin{array}{l}
(U^n)_{0,0}=\mu_n, \qquad
(U^n)_{1,1}=3\mu_n+i\sqrt{2}\mu_{n+1}-i\sqrt{2}\mu_{n-1},
\medskip \\
(U^n)_{1,0}=-(i\mu_n+\sqrt{2}\mu_{n-1}), \qquad
(U^n)_{0,1}=i\mu_n-\sqrt{2}\mu_{n+1}.
\end{array}
\]
Hence,
\[
\begin{array}{l}
u^n_{0\uparrow,0\uparrow}=
\begin{cases}
\frac{d_m}{2} & \scriptstyle n=4m,
\\
\frac{i}{\sqrt{2}} & \scriptstyle n=4m+1,
\\
-\frac{d_m}{2} & \scriptstyle n=4m+2,
\\
-\frac{i}{\sqrt{2}} & \scriptstyle n=4m+3,
\end{cases}
\qquad\kern4pt u^n_{0\downarrow,0\downarrow}=
\begin{cases}
\frac{3d_m-4}{2} & \scriptstyle n=4m,
\\
i\frac{3-2d_m}{\sqrt{2}} & \scriptstyle n=4m+1,
\\
-\frac{3d_m-4}{2} & \scriptstyle n=4m+2,
\\
-i\frac{3-d_m-d_{m+1}}{\sqrt{2}} & \scriptstyle n=4m+3,
\end{cases}
\medskip \\
u^n_{0\uparrow,0\downarrow}=
\begin{cases}
-i\frac{2-d_m}{2} & \scriptstyle n=4m,
\\
\frac{d_m-1}{\sqrt{2}} & \scriptstyle n=4m+1,
\\
i\frac{2-d_m}{2} & \scriptstyle n=4m+2,
\\
-\frac{d_{m+1}-1}{\sqrt{2}} & \scriptstyle n=4m+3,
\end{cases}
\qquad u^n_{0\downarrow,0\uparrow}=
\begin{cases}
i\frac{2-d_m}{2} & \scriptstyle n=4m,
\\
-\frac{d_m-1}{\sqrt{2}} & \scriptstyle n=4m+1,
\\
-i\frac{2-d_m}{2} & \scriptstyle n=4m+2,
\\
\frac{d_m-1}{\sqrt{2}} & \scriptstyle n=4m+3,
\end{cases}
\end{array}
\]
except for $u^1_{0\downarrow,0\downarrow}=0$ and
$u^1_{0\downarrow,0\uparrow}=-1/\sqrt{2}$.

The probability amplitudes of this example are quite different from those
of the Hadamard coin on the non-negative integers. They also show
quite a different behaviour when compared to any unbiased QRW on the
integers, including the case of the same coin (\ref{Hmod}). In particular,
if $n\ge2$, this example has no forbidden $n$-step transitions between the
pure spin states at site 0. This is due to the inequality $d_2 < d_n <
d_1$ for $n\ge3$, which is a consequence of (\ref{dn}) and the fact that
$|\hat c_n|$ is decreasing and $\hat c_n$ has alternating signs for $n\ge1$.

\section{Asymptotics of QRWs} \label{ASY}

In Sections \ref{GEN} and \ref{CON-HALF} we have seen that the transition
matrix of any QRW (on $\mathbb{Z}\ge0$ or $\mathbb{Z}$) with non trivial
distinct coins has an associated (scalar or $2\times2$-matrix valued)
measure. As in the previous section, the corresponding Carath\'{e}odory
function allows us to compute the moments of the measure which, with the
aid of the related orthonormal Laurent polynomials, provide through the
KMcG formula the amplitudes of the $n$-step transitions for any value of
$n$. This opens the possibility of studying the asymptotic behaviour of
such amplitudes when $n$ goes to infinity.

Several different authors have obtained specific asymptotic results,
mainly in the case of the Hadamard walk, by using different methods.
For a very good account, see \cite{Ko}. We have not attempted any
comparison between our rather general results and the many detailed
results in the literature. The results in this section are given to
indicate how our method could be used for similar purposes.

For instance, in the case of the Hadamard coin on the non-negative
integers, as well as for any unbiased QRW on the integers, the $n^{th}$
moment has zero limit when $n$ goes to infinity because the coefficients
$c_n$ in (\ref{cn}) satisfy $\lim_{n\to\infty}c_n=0$. Therefore,
$\lim_{n\to\infty}u^n_{0\uparrow,0\uparrow} = 0$ in these cases.

On the other hand, the unbiased QRW on the non-negative integers with coin
(\ref{Hmod}) gives moments $\mu_n$ satisfying
$\lim_{n\to\infty}|\mu_n|=1/\sqrt{2}$ because, from (\ref{dn}),
\[
\lim_{n\to\infty}d_n=\sum_{k=0}^\infty\hat
c_k=\sqrt{1+z}\big|_{z=1}=\sqrt{2}.
\]
Hence, the probabilities for returning to the spin up 0-state, if the
system was originally in such a state, converge to a non-zero limit, i.e.,
$\lim_{n\to\infty}|u^n_{0\uparrow,0\uparrow}|^2=1/2$. Indeed, although the
probability amplitudes $u^n_{0\uparrow,0\uparrow}=\mu_n$ do not converge
in this example, the quantities $i^{-n}\mu_n$ actually converge to
$1/\sqrt{2}$ when $n$ goes to infinity.

The moments are given by only a few of the coefficients of the powers of
the transition matrix $U$: the coefficient $(0,0)$ of $U^n$ is the scalar
moment $\mu_n$ in the case of the non-negative integers, while the
coefficients $(0,0)$, $(0,-1)$, $(-1,0)$ and $(-1,-1)$ of $U^n$ provide
the matrix moment $\bs{\mu}_n$ for a QRW on the integers. However, as we
saw in the previous section, the rest of the transition amplitudes can be
calculated in terms of the moments using the KMcG formulas. In fact,
according to Sections \ref{GEN} and \ref{CON-HALF}, this should be
possible, not only for QRWs with a constant coin, but for any QRW with non
trivial distinct coins. As a consequence, for all these kinds of QRWs, the
asymptotic behaviour of the moments controls the asymptotic behaviour of
the powers of the transition matrix. This is the idea behind the result
given in the following proposition.

To get a better understanding of what is coming, we start with some
remarks. Looking at our examples, we have to deal with the situation
where the $n^{th}$ moment multiplied by some phase $e^{-i\theta_n}$
is convergent. As we will see, this condition is connected with the
situation where $e^{-i\theta_n}(U^n)_{i,j}$ is convergent for any
$i,j$. Since the sequence $U^n$ is uniformly bounded because $U^n$
is unitary for any $n$, the existence of
$\lim_{n\to\infty}e^{-i\theta_n}(U^n)_{i,j}$ for any $i,j$ is
equivalent to saying that $\lim_{n\to\infty} \psi e^{-i\theta_n} U^n
\eta^\dag$ exists for any row vectors $\psi,\eta \in
L^2(\mathcal{I}\times\mathcal{S})$, which defines the familiar weak
convergence of operators, see \cite[Chapter III]{Ka76}.

The weak limit $U^\infty=(U^\infty_{i,j})$,
$U^\infty_{i,j}=\lim_{n\to\infty}e^{-i\theta_n}(U^n)_{i,j}$, when it exists,
defines an operator on $L^2(\mathcal{I}\times\mathcal{S})$ which provides
the asymptotic behaviour for $n$ going to infinity of the $n$-step
transition amplitude between any two states $\psi,\eta$ because
\[
\lim_{n\to\infty}\left(\psi U^n \eta^\dag - e^{i\theta_n} \psi U^\infty
\eta^\dag\right)=0.
\]
Although $U^\infty$ does not inherit in general the unitarity of $U^n$, it
has a norm not greater than one because $U^n$ does so for any $n$.

\begin{proposition} \label{ASY-MOM}

Let $U$ be the transition matrix of a QRW on the integers or the
non-negative integers with arbitrary non trivial distinct coins.
Concerning the asymptotic behaviour as $n$ goes to infinity of $U^n$ and
the $n^{th}$ moment of the related orthogonality measure, we have the
following results:

\begin{enumerate}

\item $U^n$ converges weakly to zero if and only if
the $n^{th}$ moment converges to zero.

\item For any sequence $e^{i\theta_n}$ of phases,
$e^{-i\theta_n}U^n$ has a non null weak limit if and only if the $n^{th}$
moment multiplied by $e^{-i\theta_n}$ converges to a non null limit and
$\lim_{n\to\infty}e^{i(\theta_{n+1}-\theta_n)}$ exists.

\end{enumerate}

\end{proposition}

\begin{proof}
Consider a QRW on the integers. Let $\bs{U}$ be the $2\times2$-block
five-diagonal matrix obtained by performing the folding (\ref{FOLD-Matr}) on
$U$. Then, $U^n$ converges weakly to zero if and only if
$\bs{U}^n$ does so.

The KMcG formula obtained in Section \ref{GEN} states that the
$2\times2$-blocks of $\bs{U}^n$ are given by
\[
(\bs{U}^n)_{i,j} = \int_\mathbb{T} z^n \bs{X}_i(z) d\bs{\mu}(z) \bs{X}_j(z)^\dag,
\]
where $\bs{X}_i$ are the corresponding orthonormal Laurent polynomials. If
$\bs{X}_i(z)=\sum_{k=p_i}^{q_i}A_{i,k}z^k$ with $2\times2$ matrix
coefficients $A_{i,k}$, then
\[
(\bs{U}^n)_{i,j} =
\sum_{k=p_i}^{q_i} \sum_{l=p_j}^{q_j} A_{i,k} \bs{\mu}_{n+k-l} A_{j,l}^\dag.
\]
This equality implies that $\lim_{n\to\infty}(\bs{U}^n)_{i,j}=\bs{0}$ for
any $i,j$ exactly when
$\lim_{n\to\infty}\bs{\mu}_n=\lim_{n\to\infty}(\bs{U}^n)_{0,0}=\bs{0}$,
which proves (1).

Concerning (2), notice that $e^{-i\theta_n}U^n$ and
$e^{-i\theta_n}\bs{U}^n$ have a non null weak limit simultaneously.
Suppose that both $\lim_{n\to\infty}e^{-i\theta_n}\bs{\mu}_n$ and
$\lim_{n\to\infty}e^{i(\theta_{n+1}-\theta_n)}$ exist. The above equality
gives
\[
e^{-i\theta_n}(\bs{U}^n)_{i,j} =
\sum_{k=p_i}^{q_i} \sum_{l=p_j}^{q_j} e^{i(\theta_{n+k-l}-\theta_n)} A_{i,k} e^{-i\theta_{n+k-l}} \bs{\mu}_{n+k-l} A_{j,l}^\dag.
\]
If $\lim_{n\to\infty}e^{i(\theta_{n+1}-\theta_n)}=z_0$, then
$\lim_{n\to\infty}e^{i(\theta_{n+k}-\theta_n)}=z_0^k$. Thus, the previous
identity shows that $\lim_{n\to\infty}e^{-i\theta_n}(\bs{U}^n)_{i,j}$
exists for any $i,j$. Moreover, if
$\lim_{n\to\infty}e^{-i\theta_n}\bs{\mu}_n\ne\bs{0}$, then
$\lim_{n\to\infty}e^{-i\theta_n}(\bs{U}^n)_{i,j}\ne\bs{0}$ at least for
$i=j=0$.

Conversely, assume that $\lim_{n\to\infty}e^{-i\theta_n}(\bs{U}^n)_{i,j}$
exists for any $i,j$ and is non null for some $i,j$. Then,
$\lim_{n\to\infty}e^{-i\theta_n}\bs{\mu}_n=\lim_{n\to\infty}e^{-i\theta_n}(\bs{U}^n)_{0,0}$
exists and must be non null because otherwise
$\lim_{n\to\infty}e^{-i\theta_n}(\bs{U}^n)_{i,j}=\bs{0}$ for any $i,j$ due
to (1). Denote by $U^\infty$ the (non null) weak limit of
$e^{-i\theta_n}U^n$. Taking weak limits in
$Ue^{-i\theta_n}U^n=e^{i(\theta_{n+1}-\theta_n)}e^{-i\theta_{n+1}}U^{n+1}$
we obtain
$UU^\infty=\lim_{n\to\infty}e^{i(\theta_{n+1}-\theta_n)}U^\infty$, thus
$e^{i(\theta_{n+1}-\theta_n)}$ must converge because $U^\infty$ is not
null.

The proof for a QRW on the non-negative integers is completely analogous
and even simpler because we do not need the folding.
\end{proof}

As a consequence of the previous results, the transition matrix of any
unbiased QRW on the integers converges weakly to zero, which means that
the amplitude of the $n$-step transition between any two (finite or
infinite) superposition of pure states converges to zero as $n$ goes to
infinity. This is also true for the Hadamard coin on the non-negative
integers. On the contrary, the transition matrix $U$ for the unbiased QRW
on the non-negative integers with coin (\ref{Hmod}) should be such that
$i^{-n}U^n$ converges weakly to some non vanishing weak limit $U^\infty$.
Indeed, we can compute such a weak limit with the aid of the following
result.

\begin{proposition} \label{ASY-MASS}

Let $U$ be the transition matrix of a QRW on the non-negative integers
with arbitrary non trivial distinct coins. If, for a sequence
$e^{i\theta_n}$ of phases, $e^{-i\theta_n}U^n$ has a non null weak limit
$U^\infty$, the related orthogonality measure $\mu$ has a mass point $z_0$
to which $e^{i(\theta_{n+1}-\theta_n)}$ converges and
\[
U^\infty = \mu_\infty X(z_0) X(z_0)^\dag,
\qquad
\mu_\infty=\lim_{n\to\infty}e^{-i\theta_n}\mu_n,
\]
with $X=(1,X_1,X_2\dots)^T$ the associated column vector of orthonormal
Laurent polynomials and $\mu_n$ the moments of $\mu$. Furthermore,
\[
\lim_{n\to\infty} e^{-i\theta_n}z_0^n=\frac{\mu_\infty}{\mu(\{z_0\})},
\]
so that
$\lim_{n\to\infty}z_0^{-n}\mu_n=\mu(\{z_0\})$ and $z_0^{-n}U^n$ converges
weakly to
\[
\mu(\{z_0\})X(z_0)X(z_0)^\dag = \frac{1}{\|X(z_0)\|^2}\,X(z_0)X(z_0)^\dag,
\]
which is the orthogonal projection onto the eigenspace of $U$ associated
with the eigenvalue $z_0$.

\end{proposition}

\begin{proof}
Suppose that $e^{-i\theta_n}U^n$ converges weakly to a non null limit
$U^\infty$. From Proposition \ref{ASY-MOM} we know that
$\lim_{n\to\infty}e^{-i\theta_n}\mu_n=\mu_\infty\neq0$ and
$\lim_{n\to\infty}e^{i(\theta_{n+1}-\theta_n)}=z_0\in\mathbb{T}$. The
arguments at the end of the proof of such proposition yield the identity
$UU^\infty=z_0U^\infty$. Thus, the non null columns of $U^\infty$ must be
eigenvectors of $U$ with eigenvalue $z_0$.

Let $\mathcal{C}$ be the CMV matrix related to $U$, and let $x$ be the
corresponding column vector of orthonormal Laurent polynomials. Bearing in
mind that $U=\Lambda\mathcal{C}\Lambda^\dag$ and $X=\Lambda x$ with
$\Lambda$ unitary diagonal, the comments in Section \ref{CMV} show that
$z_0$ must be a mass point of $\mu$ with a mass given by
$\mu(\{z_0\})=1/\|x(z_0)\|^2=1/\|X(z_0)\|^2$. Moreover, the eigenvectors
of $U$ with eigenvalue $z_0$ must be spanned by $X(z_0)$, so the columns
of $U^\infty$ should be proportional to $X(z_0)$, i.e.,
\[
U^\infty = X(z_0)Y, \qquad Y=(Y_0,Y_1,\dots) \in L^2(\mathbb{Z}\ge0).
\]
Notice that
$Y_0=U^\infty_{0,0}=\lim_{n\to\infty}e^{-i\theta_n}(U^n)_{0,0}=\mu_\infty$.

On the other hand, $e^{-i\theta_n}(U^T)^n$ converges weakly to
$(U^\infty)^T$. The unitarity of $U$ implies that $z_0$ must be an
eigenvalue of $U^T$ too, and the corresponding eigenvectors must be
spanned by $\overline{X(z_0)}$. Therefore, similar arguments to the
previous ones show that
\[
(U^\infty)^T = \overline{X(z_0)}Z, \qquad Z=(Z_0,Z_1,\dots) \in L^2(\mathbb{Z}\ge0),
\]
with $Z_0=\mu_\infty$.

Therefore, the matrix $\overline{X(z_0)}Z$ must be equal to
$Y^TX(z_0)^T$. Identifying the first column of both matrices gives
$Y^T=Z_0\overline{X(z_0)}=\mu_\infty\overline{X(z_0)}$, hence
$U^\infty = \mu_\infty X(z_0)X(z_0)^\dag$.

From $U^\infty = \mu_\infty X(z_0)X(z_0)^\dag$ we obtain
\[
\begin{aligned}
\mu_\infty \|X(z_0)\|^4 & = X(z_0)^\dag U^\infty X(z_0) =
\lim_{n\to\infty} X(z_0)^\dag e^{-i\theta_n}U^n X(z_0) =
\\
& = \lim_{n\to\infty} e^{-i\theta_n}z_0^n \|X(z_0)\|^2,
\end{aligned}
\]
which proves that $\lim_{n\to\infty} e^{-i\theta_n}z_0^n = \mu_\infty
\|X(z_0)\|^2 = \mu_\infty/\mu(\{z_0\})$. The rest of the identities follow
easily from this equality.
\end{proof}

The second part of Proposition \ref{ASY-MASS} asserts that, when
$e^{-i\theta_n}U^n$ has a non null weak limit, we can suppose without loss
of generality that $e^{i\theta_n}$ is $z_0^n$, with $z_0$ the mass point
of $\mu$ to which $e^{i(\theta_{n+1}-\theta_n)}$ must converge and, then,
$\mu_\infty$ becomes $\mu(\{z_0\})$.

For instance, in the case of the unbiased QRW on the non-negative integers
with coin (\ref{Hmod}), $\lim_{n\to\infty}i^{-n}\mu_n=1/\sqrt{2}$ is the
mass of the only mass point $z_0=i$ of the corresponding measure.
According to Proposition \ref{ASY-MASS}, the weak limit $U^\infty$ of
$i^{-n}U^n$ is
\[
U^\infty = \frac{1}{\sqrt{2}}\,X(i)X(i)^\dag, \qquad
X=(1,X_1,X_2,\dots)^T,
\]
where $X_k$ are the associated orthonormal Laurent polynomials given in
Section \ref{HAD}, so
\[
\begin{array}{l}
X_{2j-1}(i)=i^j\{U_j(\sqrt{2})-(\sqrt{2}+1)U_{j-1}(\sqrt{2})\},
\smallskip \\
X_{2j}(i)=(-1)^j\overline{X_{2j-1}(i)}=X_{2j-1}(i).
\end{array}
\]
Recurrence (\ref{eq2}) for the Chebyshev polynomials $U_j$ implies that
$U_j(\sqrt{2})=2\sqrt{2}U_{j-1}(\sqrt{2})-U_{j-2}(\sqrt{2})$, hence
\[
\begin{aligned}
X_{2j}(i) & = i^j\{(\sqrt{2}-1)U_{j-1}(\sqrt{2})-U_{j-2}(\sqrt{2})\} =
\\
& = i^j(\sqrt{2}-1)\{U_{j-1}(\sqrt{2})-(\sqrt{2}+1)U_{j-2}(\sqrt{2})\} =
\\
& = i(\sqrt{2}-1)X_{2j-2}(i).
\end{aligned}
\]
Therefore, $X_{2j-1}(i)=X_{2j}(i)=(i(\sqrt{2}-1))^j$ and
\[
U^\infty_{2j-1,2k-1}=U^\infty_{2j-1,2k}=U^\infty_{2j-1,2k}=U^\infty_{2j,2k}=
\frac{i^{j-k}}{\sqrt{2}}(\sqrt{2}-1)^{j+k}.
\]
This allows us to compute the asymptotic transition amplitude between any
two states. In particular, denoting $\lim_{n\to\infty}(a_n/b_n)=1$ by
$a_n\sim_nb_n$,
\[
\begin{array}{l}
\displaystyle
u^n_{j\uparrow,k\uparrow}\sim_n\frac{i^{n+j-k}}{\sqrt{2}}(\sqrt{2}-1)^{j+k},
\kern37pt
u^n_{j\downarrow,k\downarrow}\sim_n\frac{i^{n+j-k}}{\sqrt{2}}(\sqrt{2}-1)^{j+k+2},
\medskip \\ \displaystyle
u^n_{j\uparrow,k\downarrow}\sim_n\frac{i^{n+j-k-1}}{\sqrt{2}}(\sqrt{2}-1)^{j+k+1},
\kern15pt
u^n_{j\downarrow,k\uparrow}\sim_n\frac{i^{n+j-k+1}}{\sqrt{2}}(\sqrt{2}-1)^{j+k+1}.
\end{array}
\]

These asymptotic results give further indications of the different
probabilistic behaviour of an unbiased QRW when considered on the
non-negative integers or in all the integers.

\section{Recurrence properties of QRWs} \label{REC}

For a classical random walk there is an important notion that goes back at
least to G. Polya, see \cite{Po}. We say that a state is recurrent if,
having started there at the initial time, one returns to it with
probability one. Otherwise we say that the state is transient. For a
so-called irreducible chain either all states are recurrent or they are all transient.
A recurrent state is called positive recurrent if the expected value for
the time of (first) return to it is finite. When dealing with a
birth-and-death process on the non-negative integers the corresponding
orthogonality measure $dm(x)$ has support in the interval $[-1,1]$ and
plays an important role in studying these notions.

For instance, the process
is recurrent exactly when
\begin{equation} \label{C-REC-0}
\int_{-1}^1 \frac{dm(x)}{1-x} = \infty.
\end{equation}
Notice that this integral is the sum of all moments of the measure $dm(x)$
and that the $n^{th}$ moment is the probability of going from the state $0$
to itself in $n$ steps.

The process returns to the origin in a finite expected time when the
measure has a mass at $x=1$. The existence of
\[
\lim_{n \to \infty} (P^n)_{i,j}
\]
is equivalent to $dm(x)$ having no mass at $x=-1$.  If this is the case
this limit is positive exactly when $dm(x)$ has some mass at $x = 1$.

In the classical case one gets a lot of milage out of the generating
function $S(z)$ of the moments of $dm(x)$ given by
\[
S(z)=\int_{-1}^1 \frac {dm(x)}{1-x z}.
\]

In particular the generating function $G(z)$ of the sequence $g_n$ giving
the probability of a first return to the origin in $n$ steps
\[
G(z) = \sum_{n=0}^{\infty} z^ng_n
\]
is related to $S(z)$ by
\[
G(z) = 1 - \frac {1}{S(z)}.
\]

Therefore we have that $G(1) = 1$  (indicating that one returns to
state $0$ with probability one) exactly when $S(1)$ is infinite as
noticed above. This relation allows us to compute the expected time
to return to state $0$. This expected value is given by $G'(1).$

We are confident that these ideas and results should have a natural
translation to the quantum case using the tools provided by the
orthonormal Laurent polynomials on the unit circle. Indeed, an approach to
the asymptotics of the powers of the transition matrix using such a
machinery has been presented in the previous section.

We will see now that the notion of quantum recurrence can be
described nicely in terms of the Carath\'{e}odory function introduced
earlier, which will play in the quantum case a similar role to the
generating function $S(z)$ of the moments for a classical random
walk. Nevertheless, the condition for the characterization of
quantum recurrence will be somewhat different from (\ref{C-REC-0}).

The study of the recurrence properties in the quantum case raises special
issues because a quantum measurement destroys the initial evolution since
the system collapses into a pure state when a measurement is performed.
Therefore, the notion of a return to a given state {\bf for the first
time} in a certain number of steps has to be interpreted with care in the
quantum case. Such an analysis in terms of a specific measurement scheme,
involving an ensemble of identically prepared QRWs, has been proposed
recently, see \cite{SKJ}.

The bottom line of this analysis is that, in the quantum case, the
recurrence of a state is characterized by the divergence of the series of
probabilities to return to such a state in $n$ steps. After the
modifications which are necessary to make sense of the notion of
recurrence in the quantum case, this result is completely analogous to the
classical one. Its importance lies on the fact that the notion of a return
to a state in a certain number of steps is completely meaningful in the
quantum case and its probability can be computed using the transition
matrix of the QRW.

More precisely, following the interpretation of the quantum recurrence
given in \cite{SKJ}, a state $\psi$ of a QRW with transition matrix $U$ is
recurrent exactly when
\begin{equation} \label{REC-pn}
\sum_{n=1}^\infty p_n(\psi) = \infty, \qquad p_n(\psi)=|\psi U^n \psi^\dag|^2,
\end{equation}
where $p_n(\psi)$ stands for the probability to return to the state
$\psi$ in $n$ steps.

Consider a QRW on the non-negative integers with non trivial distinct
coins. The recurrence of the state numbered as 0, i.e., the spin up at
site 0, is characterized by the divergence of
$\sum_{n=1}^\infty|(U^n)_{0,0}|^2=\sum_{n=1}^\infty|\mu_n|^2$. From the
McLaurin series (\ref{MOM}) of the related Carath\'{e}odory function $F(z)$ we
obtain
\[
\int_0^{2\pi} |F(e^{i\theta})|^2 \frac{d\theta}{2\pi} =
1 + 2\sum_{n=1}^\infty|\mu_n|^2,
\]
where $F(e^{i\theta})=\lim_{r\uparrow1}F(re^{i\theta})$, which exists for
Lebesgue almost every $\theta\in[0,2\pi)$, see \cite[Chapter 17]{Ru74}.
Therefore, $|0\>\otimes\uk$ is recurrent exactly when the radial limit of
$F(z)$ does not lie on $L^2_\frac{d\theta}{2\pi}(\mathbb{T})$.

The generalization of this result to an arbitrary state, as well as to
QRWs on the integers, is the purpose of the proposition below. In what
follows we write $F(z)=F(z,d\mu)$ when we need to make explicit the
measure corresponding to a Carath\'{e}odory function, and similarly for matrix
valued Carath\'{e}odory functions. Besides, we will assume that any scalar or
matrix valued Carath\'{e}odory function is radially extended Lebesgue almost
everywhere on the unit circle (see \cite{DaPuSi08} for the matrix case).

\begin{proposition} \label{Q-REC}

Consider a QRW with non trivial distinct coins on
$\mathcal{I}=\mathbb{Z}\text{ or }\mathbb{Z}\ge0$.

\begin{enumerate}

\item If $\mathcal{I}=\mathbb{Z}\ge0$, let $\mu$ be the related orthogonality measure and let
us number the states {\small
$|0\>\otimes\uk,|0\>\otimes\dk,|1\>\otimes\uk,|1\>\otimes\dk,\dots$} as
$|0\>,|1\>,|2\>,|3\>,|4\>,\dots$. A state
$|\Psi\>=\sum_{k=0}^\infty\psi_k|k\>$ is transient if and only if
\[
F(z,|f|^2d{\mu}) \in L^2_{\frac{d\theta}{2\pi}}(\mathbb{T}),
\qquad
f = \sum_{k=0}^\infty \psi_kX_k,
\]
where $X_k$ are the corresponding orthonormal Laurent polynomials.

\item If $\mathcal{I}=\mathbb{Z}$, let $\bs{\mu}$ be the related orthogonality matrix measure
and let us number the states {\small
$|0\>\otimes\uk,|\!-\!1\>\otimes\dk,|\!-\!1\>\otimes\uk,|0\>\otimes\dk,\dots$}
as $|0\>,|1\>,|2\>,|3\>,|4\>,\dots$. A state
$|\Psi\>=\sum_{k=0}^\infty\psi_k|k\>$ is transient if and only if
\[
F(z,\bs{f}d\bs{\mu}\bs{f}^\dag) \in L^2_{\frac{d\theta}{2\pi}}(\mathbb{T}),
\qquad
\bs{f} = \sum_{k=0}^\infty (\psi_{2k},\psi_{2k+1})\bs{X}_k,
\]
where $\bs{X}_k$ are the corresponding matrix orthonormal Laurent polynomials.

\end{enumerate}

\end{proposition}

\begin{proof}
We will only prove (2) since the proof of (1) is similar and simpler.
Consider a QRW on the integers with non trivial distinct coins. Let
$\bs{U}$ be the semi-infinite transition matrix obtained performing the
folding (\ref{FOLD-Matr}) on the doubly infinite transition matrix $U$.
The recurrence of the state $|\Psi\>$ is equivalent to the divergence of
$\sum_{n=0}^\infty p_n(\psi)$, where $p_n(\psi)=|\psi\bs{U}^n\psi^\dag|^2$
and $\psi=(\psi_0,\psi_1,\dots)$ is the wave function of $|\Psi\>$
corresponding to the same folding.

Denoting $\bs{\psi}_k=(\psi_{2k},\psi_{2k+1})$, the KMcG formula permits
us to write
\[
\begin{aligned}
\psi\bs{U}^n\psi^\dag & =
\sum_{j,k=0}^\infty \bs{\psi}_j (\bs{U}^n)_{j,k} \bs{\psi}_k^\dag =
\sum_{j,k=0}^\infty \int_\mathbb{T} z^n \bs{\psi}_j \bs{X}_j(z) d\bs{\mu}(z) \bs{X}_k(z) \bs{\psi}_k^\dag =
\\
& = \int_\mathbb{T} z^n \bs{f}(z) d\bs{\mu}(z) \bs{f}(z)^\dag,
\qquad\qquad \bs{f} = \sum_{k=0}^\infty \bs{\psi}_k \bs{X}_k.
\end{aligned}
\]
The above equality identifies $\psi\bs{U}^n\psi^\dag$ as the $n^{th}$
moment of the scalar measure $\bs{f} d\bs{\mu} \bs{f}^\dag$. Therefore,
the same arguments given before the proposition show that the divergence
of $\sum_{n=0}^\infty |\psi\bs{U}^n\psi^\dag|^2$ is equivalent to
$F(z,\bs{f}d\bs{\mu}\bs{f}^\dag) \notin
L^2_\frac{d\theta}{2\pi}(\mathbb{T})$.
\end{proof}

Notice that this proposition associates with any state a scalar function
$f \in L^2_\mu(\mathbb{T})$ or a 2-dimensional vector valued function
$\bs{f} \in L^2_{\bs\mu}(\mathbb{T})$, depending on whether the QRW is on
$\mathbb{Z}\ge0$ or $\mathbb{Z}$, so that $F(z,|f|^2d\mu)$ and
$F(z,\bs{f}d\bs{\mu}\bs{f}^\dag)$ are both scalar Carath\'{e}odory functions.
We will refer to $f$ and $\bs{f}$ as the functions associated with the
corresponding state.

The condition (\ref{C-REC-0}) for the classical recurrence at the origin
depends only on the behaviour of generating function $S(z)$ of the moments
at $z=1$. In contrast, the characterization of the quantum recurrence in
terms of Carath\'{e}odory functions has to do with their global behaviour on
the whole unit circle.

Proposition \ref{Q-REC} has the drawback that it is not given in terms of
the Carath\'{e}odory function of the original measure associated with the QRW,
which is the one that we directly know, but in terms of the Carath\'{e}odory
function of a modification of such a measure. The following result shows
that this problem can be overcome, at least when analyzing the recurrence
of a local state, i.e., a state which is a superposition of a finite
number of pure states.

Notice that the function associated with any local state is a scalar
Laurent polynomial $f$ in the case $\mathbb{Z}\ge0$, or a 2-dimensional
vector valued Laurent polynomial $\bs{f}$ in the case $\mathbb{Z}$.

\begin{proposition} \label{F-Ff}

For any positive definite $d \times d$ matrix valued measure $\bs{\mu}$
and any $d$-dimensional vector valued Laurent polynomial $\bs{f}$
\[
F(z,\bs{f}d\bs{\mu}\bs{f}^\dag) \in L^2_\frac{d\theta}{2\pi}
\Leftrightarrow \bs{f}(z)\bs{F}(z,d\bs{\mu})\bs{f}(z)^\dag \in
L^2_\frac{d\theta}{2\pi}.
\]

\end{proposition}

\begin{proof}
If $\bs{f}=\sum_{k=p}^q\bs{a}_kz^k$, $\bs{a}_k\in\mathbb{C}^d$, then
\[
F(z,\bs{f}d\bs{\mu}\bs{f}^\dag) = \sum_{j,k=p}^q \bs{a}_j \bs{F}_{j-k}(z) \bs{a}_k^\dag,
\qquad
\bs{F}_k(z)=\int_\mathbb{T}\frac{t+z}{t-z}\,t^k d\bs{\mu}(t).
\]

Writing
\[
\frac{t+z}{t-z}\,t^k = (t+z)t^{k-1} + z\,\frac{t+z}{t-z}\,t^{k-1}
\]
we find that
\[
\bs{F}_k(z) = \bs{\mu}_k + \bs{\mu}_{k-1} z + z \bs{F}_{k-1}(z).
\]
Bearing in mind that $\bs{F}_0(z)=\bs{F}(z,d\bs{\mu})$, the iteration of
the above equality yields $\bs{F}_k(z)=\bs{L}_k(z)+z^k\bs{F}(z,d\bs{\mu})$
for some matrix valued Laurent polynomial $\bs{L}_k$.

In consequence, $F(z,\bs{f}d\bs{\mu}\bs{f}^\dag) = L(z) +
\bs{f}(z)\bs{F}(z,d\bs{\mu})\bs{f}(z)^\dag$ for some scalar Laurent
polynomial $L$. This relation proves the proposition.
\end{proof}

Although this proposition holds for matrix measures of arbitrary dimension
$d$, we will use it only for $d=1,2,$ which are the cases related to QRWs
on $\mathbb{Z}\ge0$ and $\mathbb{Z}$. Combining this result and
Proposition \ref{Q-REC} we see that the local transient states can be
characterized by $|f|^2F(z,d{\mu}) \in
L^2_{\frac{d\theta}{2\pi}}(\mathbb{T})$ in $\mathbb{Z}\ge0$ and
$\bs{f}(z)\bs{F}(z,d\bs{\mu})\bs{f}(z)^\dag \in
L^2_{\frac{d\theta}{2\pi}}(\mathbb{T})$ in $\mathbb{Z}$.

These alternative conditions provide a very practical way to determine the
local transient states: they must have as an associated function a Laurent
polynomial which cancels the singularities of the Carath\'{e}odory function
which are responsible for the non integrability.

Remember that the state $|0\>\otimes\uk$ of a QRW on $\mathbb{Z}\ge0$ is
transient exactly when $F(z,d\mu) \in
L^2_\frac{d\theta}{2\pi}(\mathbb{T})$. Then, $|f(z)|^2F(z,d\mu) \in
L^2_\frac{d\theta}{2\pi}(\mathbb{T})$ for any Laurent polynomial $f$, so
we find the following direct consequence of the previous results.

\begin{corollary} \label{TRANS}

The local states of a QRW on the non-negative integers with non trivial
distinct coins are all transient if and only if the state $|0\>\otimes\uk$
is transient.

\end{corollary}

Let us apply the previous results to the analysis of the recurrence for
the examples of QRWs given in Section \ref{HAD}.

Consider first the Hadamard coin on the non-negative integers. The
corresponding Carath\'{e}odory function can be written as
\[
F(z) = \sqrt{\frac{(z+z_0)(z+\overline z_0)}{(z-z_0)(z-\overline z_0)}},
\qquad z_0=\frac{1}{\sqrt{2}}(1+i),
\]
for some choice of the square root. $|F|^2$ has two singularities on
$\mathbb{T}$: $z_0$ and $\overline z_0$. Neither is Lebesgue
integrable, so the state given by a  spin up at the origin is
recurrent.

Local transient states are characterized by an associated Laurent
polynomial $f$ such that $|f|^4|F|^2$ is Lebesgue integrable on
$\mathbb{T}$. Therefore, the local transient states are those with an
associated Laurent polynomial vanishing at $z_0$ and $\overline z_0$.

Any superposition of up and down states at site 0 has an associated
function lying in
$\text{span}\{X_0,X_1\}=\text{span}\{1,z^{-1}\}=z^{-1}\text{span}\{1,z\}$.
Such a function can not cancel both singularities of $F$, thus any
mixed spin state at the origin is recurrent.

However, transient states can appear if we consider a mixing of spin
states at sites 0 and 1. The Laurent polynomial associated with a state
$a\,|0\>\otimes\uk + b\,|0\>\otimes\dk + c\,|1\>\otimes\uk$ is in
$\text{span}\{X_0,X_1,X_2\}=\text{span}\{1,z^{-1},z^2\}=z^{-1}\text{span}\{1,z,z^2\}$,
so it can vanish at both, $z_0$ and $\overline z_0$. The Laurent
polynomial related to such a state is $a+bX_1+cX_2$, so it is transient
exactly when
\[
a+bX_1(z_0)+cX_2(z_0)=0, \qquad a+bX_1(\overline z_0)+cX_2(\overline z_0)=0.
\]

Since $X_1(z)=1-\sqrt{2}z^{-1}$ and $X_2(z)=-X_1(1/z)$ we find that the
solutions of the above equations are $a=0$ and $c=-b$. This means that the
transient states with the referred form are spanned by
\[
|0\>\otimes\dk - |1\>\otimes\uk.
\]

Following a similar reasoning, and using the form of the third
orthonormal Laurent polynomial
$X_3(z)=1+\sqrt{2}(z-z^{-1})(1-\sqrt{2}z^{-1})$, it is easy to
obtain the transient states mixing all the up and down states at
sites 0 and 1. The result is that such transient states are those
lying in the span of
\[
|0\>\otimes\dk - |1\>\otimes\uk, \qquad |0\>\otimes\uk + |1\>\otimes\dk.
\]

Let us see what happens if we change the Hadamard coin by another
equiprobable coin like (\ref{Hmod}). Then, the Carath\'{e}odory function
\[
F(z) = \frac{\sqrt{1+z^4}-i\sqrt{2}z}{z^2+1}
\]
has a single non integrable singularity at $i$ because $-i$ is a removable
one. Thus the spin up at the origin is recurrent once again.

However, in contrast with the Hadamard coin, this QRW has transient states
at the origin. Such transient states $a\,|0\>\otimes\uk+b\,|0\>\otimes\dk$
have an associated Laurent polynomial $a+bX_1$ which must vanish at $i$.
Since $X_1(z)=-(\sqrt{2}z^{-1}+i)$ we find that $a+bX_1(i)=0$ is solved by
$b=i(1+\sqrt{2})a$, which shows that the states spanned by
\[
|0\>\otimes\uk + i(1+\sqrt{2})\,|0\>\otimes\dk
\]
are transient.

We can also look for the transient states mixing the spin states at sites
0 and 1. Using the fact that $X_2(z)=-\overline{X_1(1/\overline z)}$ and
$X_3(z)=1+i\sqrt{2}(z-z^{-1})(i\sqrt{2}z^{-1}-1)$ we obtain a
3-dimensional subspace of transient states $a\,|0\>\otimes\uk +
b\,|0\>\otimes\dk + c\,|1\>\otimes\uk + d\,|1\>\otimes\dk$ given by the
equation
\[
a+i(\sqrt{2}-1)(b+c)+(i(\sqrt{2}-1))^2d=0.
\]

The QRWs analyzed above are archetypical examples of QRWs on
the non-negative integers with a non trivial constant coin. They show the
two possible recurrence behaviours, if one leaves aside the singular case of a
diagonal coin, which is related to null Verblunsky parameters.

If the coin is symmetric the QRW is associated with an imaginary parameter
$a=\overline{c}_{21}e^{i\vartheta}$. Then, the expression of the
Carath\'{e}odory function $F$ given in Section \ref{CON-HALF} shows that
$|F|^2$ has two non integrable singularities on the unit circle. Indeed,
like in the Hadamard case, $F^2$ is a quotient of two coprime polynomials
of degree 2 with their roots on $\mathbb{T}$. Hence, the recurrence
properties for a (non trivial and non diagonal) symmetric coin on the
non-negative integers are qualitatively similar to those obtained for the
Hadamard one. In particular, any state at site 0 is recurrent for such a
coin.

On the contrary, a non symmetric coin is related to a parameter
$a=\overline{c}_{21}e^{i\vartheta}$ with a non null real part. The
corresponding Carath\'{e}odory function $F$ has only one non removable
singularity $z_0$ on the unit circle. More precisely,
$F(z)=F_0(z)/(z-z_0)$ with $|F_0|^2$ integrable on $\mathbb{T}$ and
$F_0(z_0)\neq0$, exactly as for the coin (\ref{Hmod}). Therefore, the
recurrence properties for the coin (\ref{Hmod}) are qualitatively the same
as for any other (non trivial and non diagonal) non symmetric coin on the
non-negative integers. For instance, these coins always have a
1-dimensional transient subspace at site 0.

Finally, consider the Hadamard coin on the integers. Just as in the case
of the Hadamard coin on the non-negative integers, the Carath\'{e}odory
function
\[
\bs{F}(z) = \frac{1}{\sqrt{1+z^4}} \, \bs{F}_0(z),
\qquad
\bs{F}_0(z) = \begin{pmatrix}
1+z^2 & \sqrt{2}z \\ \sqrt{2}z & 1+z^2
\end{pmatrix},
\]
has singularities at $z_0=(1+i)/\sqrt{2}$ and $\overline z_0$, but also at
$-z_0$ and $-\overline z_0$. Any of them can cause the non integrability
of $|\bs{f}\bs{F}\bs{f}^\dag|^2$ for a 2-dimensional vector valued Laurent
polynomial $\bs{f}$.

The local transient states are those whose associated vector Laurent
polynomial $\bs{f}$ is such that the scalar Laurent polynomial
$\bs{f}\bs{F}_0\bs{f}^\dag$ vanishes at $\pm z_0$ and $\pm\overline z_0$.
On these singularities $\bs{F}_0$ is proportional to a semidefinite matrix,
\[
\bs{F}_0(\pm z_0)=(1+i)\begin{pmatrix} 1 & \pm1 \\ \pm1 & 1 \end{pmatrix},
\qquad
\bs{F}_0(\pm\overline z_0)=(1-i)\begin{pmatrix} 1 & \pm1 \\ \pm1 & 1 \end{pmatrix},
\]
hence $\bs{f}\bs{F}_0\bs{f}^\dag$ vanishes on such points if and only if
$\bs{f}\bs{F}_0$ does so.

Any vector Laurent polynomial
$\bs{f}=(a_1,a_2)\bs{X}_0+(b_1,b_2)\bs{X}_1$ has the form
$\bs{f}(z)=z^{-1}\bs{p}(z)$ where $\bs{p}$ is a vector polynomial
with $\deg\bs{p}\le1$. Therefore, $\deg\bs{p}\bs{F}_0\le3$ and
$\bs{f}\bs{F}_0$ can not vanish on four different points. This means
that any superposition of spin states at sites $-1$ and 0 is
recurrent. Taking into account the translation invariance of the
QRW, we find that any superposition of spin states which mixes only
two contiguous sites is recurrent.

Thus, a transient state must involve sites which are not contiguous.
The simplest way to do that is to consider a vector Laurent
polynomial
$\bs{f}=(a_1,a_2)\bs{X}_0+(b_1,b_2)\bs{X}_1+(c_1,c_2)\bs{X}_2$ which
corresponds to a state {\small $c_2\,|-2\>\otimes\dk +
b_1\,|-1\>\otimes\uk + a_2\,|-1\>\otimes\dk + a_1\,|0\>\otimes\uk +
b_2\,|0\>\otimes\dk + c_1\,|1\>\otimes\uk$}. Using the expressions
\[
\bs{X}_1(z) = \begin{pmatrix} \sqrt{2}z^{-1} & -1 \\ 1 & -\sqrt{2}z^{-1} \end{pmatrix},
\qquad
\bs{X}_2(z) = \bs{X}_1(1/z),
\]
the conditions $\bs{f}(\pm z_0)\bs{F}_0(\pm z_0)=\bs{f}(\pm\overline
z_0)\bs{F}_0(\pm\overline z_0)=0$ become
\[
\begin{array}{l}
(a_1,a_2) \begin{pmatrix} 1 \\ \pm1 \end{pmatrix} +
i(b_1,b_2) \begin{pmatrix} \mp1 \\ 1 \end{pmatrix} +
i(c_1,c_2) \begin{pmatrix} \pm1 \\ -1 \end{pmatrix} = 0,
\medskip \\
(a_1,a_2) \begin{pmatrix} 1 \\ \pm1 \end{pmatrix} +
i(b_1,b_2) \begin{pmatrix} \pm1 \\ -1 \end{pmatrix} +
i(c_1,c_2) \begin{pmatrix} \mp1 \\ 1 \end{pmatrix} = 0,
\end{array}
\]
which have the solutions $a_1=a_2=0$, $b_1=c_1$ and $b_2=c_2$. That is,
the transient states obtained are spanned by
\[
|-2\>\otimes\dk + |0\>\otimes\dk,
\qquad
|-1\>\otimes\uk + |1\>\otimes\uk.
\]
Then, the translation invariance permits us to identify as transient
subspaces all those spanned by states with the form
\[
|k\>\otimes\dk + |k+2\>\otimes\dk,
\qquad
|k+1\>\otimes\uk + |k+3\>\otimes\uk.
\]

These kinds of results are not specific of the Hadamard QRW, but similar
recurrence properties hold for any non trivial and non
diagonal constant coin on the integers. Such recurrence
properties are a consequence of the general expression for the
Carath\'{e}odory function $\bs{F}$ obtained in Section \ref{CON-LINE}, which
shows that $\bs{F}$ has four singularities on the unit circle, with the
only exception being the case of null Verblunsky parameters, which
corresponds to a diagonal coin. More precisely, for an arbitrary non
trivial and non diagonal coin, $\bs{F}=(1/\sqrt{q})\bs{F}_0$ with $q$ a
scalar polynomial of degree 4 with 4 different roots on $\mathbb{T}$ and
$\bs{F}_0$ a matrix polynomial of degree 2 which is proportional to a
semidefinite non null matrix on the roots of $q$. These general results
are the only ingredients necessary to deduce recurrence properties
qualitatively similar to those ones obtained above for the Hadamard QRW on
the integers.

The fact that any state at a given site is recurrent for an unbiased
QRW on the integers was proved in \cite{SKJ}. However, the fact that
the states mixing only two consecutive sites are recurrent too, as
well as the existence of transient states involving non-contiguous
sites is new. Moreover, the comments of the previous paragraph show
that these recurrence properties also hold for any QRW with a non
trivial and non diagonal constant coin. These general results,
together with the analysis of the recurrence for QRWs on the
non-negative integers, constitute a novelty which illustrates some
of the possibilities of this new approach to QRWs.

\section{Conclusions} \label{CONC}

Classical random walks have been traditionally studied using three
different methods. Two of these have already been used in the quantum
case. In this paper we propose an approach to the study of QRWs that is
inspired by the third of these methods.

Our approach reproduces known results, but also provides new ones and new
methods of analysis, like the use of the orthonormal Laurent polynomials
to study the asymptotics or the analysis of quantum recurrence using
Carath\'{e}odory functions.

This approach can handle non translation invariant QRWs, as well as
situations where the structure of the one step transitions is richer
than the ones considered so far. We intend to study some of these
cases in the future by using CMV matrices where all the Verblunsky
parameters are allowed to be non-zero; the examples discussed here
have a simpler structure. This approach can also be adapted in a
natural way to deal with cases when the walk can go to infinity in
rather complicated networks as well as in the case of regular
networks in various dimensions.

Whereas in the classical case when dealing with an irreducible random walk
we have a simple dicotomy: either all states are recurrent or all states
are transient, we have seen here examples where the situation in the
quantum case is much more involved. This remains as an important area for
further investigation.

It is also important to note that this approach can be easily adapted to
the case when the number of degrees of freedom in our spins is arbitrary.
One should consider some of the examples in \cite{SKJ}, such as the Grover
or the Fourier ones.

Since the effective use of this approach rests on one's ability to give
concrete expressions for the orthogonal polynomials and the orthogonality
measure going along with a given CMV matrix there are two natural ways to
proceed: start with some of the examples where all the spectral data is
known,  such as those in \cite{Si04-1}, and explore the nature of the
corresponding QRW, or conversely start with some QRW of interest and try
to compute its associated orthogonality measure and polynomials. This is
what we have done in this paper.

It would be nice to look into the QRW that goes along with the
analog of the Gaussian in the circle, namely the Rogers-Szeg\H o
case. It would also be of interest to study examples where the
measure is purely discrete.

These, as well as many other questions, remain as an interesting
challenge.

\section{Appendix} \label{APP}

Let us calculate the measure $\hat\mu$ and the orthonormal Laurent
polynomials $\hat x_n$ with Verblunsky parameters $a, \, 0, \, a, \, 0, \,
a, \, 0, \, \dots$ for an arbitrary complex number $a$ with $|a|<1$. For
convenience we will omit the ``hats" in what follows.

Setting $\rho=\sqrt{1-|a|^2}$, the related CMV matrix is
\[
\mathcal{C} = \begin{pmatrix}
\begin{array}{r|rr|rr|rr|rr}
\overline{a} & 0 & \rho & & & & \\
\rho & 0 & -a & & & & \\
\hline
& \overline{a} & 0 & 0 & \rho & & \\
& \rho & 0 & 0 & -a & & \\
\hline
& & & \overline{a} & 0 & 0 & \rho \\
& & & \rho & 0 & 0 & -a & \\
\hline
& & & & & & \ddots & & \ddots
\end{array}
\end{pmatrix}.
\]

The second recurrence of (\ref{CX}), which determines the sequence $x_j$,
states that $(\mathcal{C}-z)x=0$, $x_0=1$, which can be written as
\[
\begin{pmatrix} \overline{a} & -z \\ \rho & 0 \end{pmatrix}
\begin{pmatrix} x_{2n-1} \\ x_{2n} \end{pmatrix}
+
\begin{pmatrix} 0 & \rho \\ -z & -a \end{pmatrix}
\begin{pmatrix} x_{2n+1} \\ x_{2n+2} \end{pmatrix}
= \bs{0},
\qquad
x_0=x_{-1}=1,
\]
or equivalently
\[
\begin{pmatrix} x_{2n+1} \\ x_{2n+2} \end{pmatrix}
= T \begin{pmatrix} x_{2n-1} \\ x_{2n} \end{pmatrix},
\qquad
T = \frac{1}{\rho}
\begin{pmatrix} z^{-1} & -a \\ -\overline{a} & z \end{pmatrix},
\qquad
\begin{pmatrix} x_{-1} \\ x_0 \end{pmatrix}
= \begin{pmatrix} 1 \\ 1 \end{pmatrix}.
\]

From this identity we see by induction that
$x_{2n}(z)=\overline{x_{2n-1}(1/\overline{z})}$. Using
(\ref{OP-OLP}) and (\ref{RR-OP}), this is simply a consequence of
the vanishing of the odd Verblunsky parameters. Thus we only need to
calculate $x_{2n-1}$. We  also have
\[
\begin{pmatrix} x_{2n-1} \\ x_{2n} \end{pmatrix}
= T^n  \begin{pmatrix} 1 \\ 1 \end{pmatrix}.
\]
Consequently, if $\lambda_\pm$ are the eigenvalues of $T$, there exist
coefficients $B_\pm$ independent of $n$ such that
\begin{equation} \label{OLP-EIG}
x_{2n-1}=B_+\lambda_+^n+B_-\lambda_-^n.
\end{equation}

The eigenvalues $\lambda_\pm$ of $T$ are the solutions $\lambda$ of
\begin{equation} \label{EIG}
\lambda^2-\rho^{-1}(z+z^{-1})\lambda+1=0.
\end{equation}
Therefore, $x_{2n-1}(z)$ are solutions $Y_n$ of the second order
difference equation
\[
Y_{n+1} + Y_{n-1} = 2yY_n, \qquad y=\frac{1}{2\rho}(z+z^{-1}),
\]
which is solved by $U_n(y)$, $U_n$ being the Chebysev polynomials of
second kind, given by (\ref{eq2}). Indeed, the sequences $U_n(y)$ and
$U_{n-1}(y)$ are independent solutions of this difference equation, thus,
there exist coefficients $\gamma,\delta$ independent of $n$ such that
\[
x_{2n-1} = \gamma\,U_n(y) + \delta\,U_{n-1}(y),
\]

Evaluating this identity for $x_{-1}(z)=1$ and
$x_1(z)=\rho^{-1}(z^{-1}-a)$ yields
\[
\gamma(z)=1, \qquad \delta(z)= -\rho^{-1}(z+a),
\]
giving finally
\begin{equation} \label{xn}
\begin{aligned}
x_{2n-1}(z) = U_n(y)-\rho^{-1}(z+a)U_{n-1}(y),
\\
x_{2n}(z) = U_n(y)-\rho^{-1}(z^{-1}+\overline{a})U_{n-1}(y),
\end{aligned}
\qquad
y=\frac{1}{2\rho}(z+z^{-1}).
\end{equation}
This gives the orthonormal Laurent polynomials $(x_j)_{j=0}^\infty$.

To find the corresponding orthogonality measure $\mu$ we proceed with the
calculation of the Carath\'{e}odory function $F(z)$, $|z|<1$, using
(\ref{CF-OP}). Bearing in mind (\ref{OP-OLP}) we can write
\begin{equation} \label{CF-OLP}
F(z) = \lim_{n\to\infty} \frac{\tilde{x}_{2n-1}(z)}{x_{2n-1}(z)},
\qquad |z|<1,
\end{equation}
where $\tilde{x}_j$ are the orthonormal Laurent polynomials with
Verblunsky parameters
\[
-a, \; 0, \; -a, \; 0, \; -a, \; 0, \; \dots
\]

To take the limit (\ref{CF-OLP}) we will use the expression
(\ref{OLP-EIG}) for $x_j$, and a similar one changing $a \to -a$ for
$\tilde{x}_j$. The eigenvalues $\lambda_\pm$ of $T$ are
\[
\lambda_\pm = \frac{1}{2\rho} (z+z^{-1}\pm\sqrt{(z-z^{-1})^2+4|a|^2}),
\]
where we choose the square root so that $|\lambda_+|>|\lambda_-|$ for
$0<|z|<1$ (using $\lambda_+\lambda_-=1$ and $\lambda_+ + \lambda_- =
\rho^{-1}(z+z^{-1})$ it is not difficult to see that
$|\lambda_+|\neq|\lambda_-|$ for $0<|z|<1$ ). Then, (\ref{CF-OLP}) gives
\[
F = \frac{\tilde{B}_+}{B_+},
\]
where $\tilde{B}_+$ is obtained from $B_+$ changing $a \to -a$.

Using (\ref{OLP-EIG}) for $n=0,1$ we obtain
\[
B_+ = \frac{\rho^{-1}(z^{-1}-a)-\lambda_-}{\lambda_+-\lambda_-}.
\]
Thus, the invariance of $\lambda_\pm$ under $a \to -a$ yields
\[
\begin{aligned}
& F(z) = \frac{z^{-1}+a-\rho\lambda_-}{z^{-1}-a-\rho\lambda_-} =
\frac{z-z^{-1}-\sqrt{(z-z^{-1})^2+4|a|^2}-2a}{z-z^{-1}-\sqrt{(z-z^{-1})^2+4|a|^2}+2a} =
\\
& = -\frac{\sqrt{(z-z^{-1})^2+4|a|^2}+2\re a}{z-z^{-1}+2i\im a}
= -\frac{z-z^{-1}-2i\im a}{\sqrt{(z-z^{-1})^2+4|a|^2}-2\re a}.
\end{aligned}
\]

We can obtain the weight $w(\theta)$ of the decomposition (\ref{RN})
for $\mu$ using (\ref{CF-w}) by taking the limit $r\uparrow1$ of
$\re F(z)$, $z=re^{i\theta}$. Taking into account the choice we have
made for the square root, when $|\sin\theta|\leq|a|$,
\[
\lim_{r\uparrow1} \sqrt{(z-z^{-1})^2+4|a|^2} =
\pm 2\sqrt{|a|^2-\sin^2\theta}
\quad \text{ if }
\begin{cases}
\cos\theta\geq0,
\\
\cos\theta\leq0,
\end{cases}
\]
while, for $|\sin\theta|\geq|a|$,
\[
\lim_{r\uparrow1} \sqrt{(z-z^{-1})^2+4|a|^2} =
\mp 2i \sqrt{\sin^2\theta-|a|^2}
\quad \text{ if }
\begin{cases}
\sin\theta\geq0,
\\
\sin\theta\leq0.
\end{cases}
\]
This allows one to obtain $w(\theta)=\lim_{r\uparrow1}\re F(re^{i\theta})$
which is given by
\[
w(\theta) =
\frac{\sqrt{\sin^2\theta-|a|^2}}{|\sin\theta+\im a|},
\qquad |\sin\theta|\geq|a|,
\]
and zero otherwise. Equivalently,
\[
w(\theta) = \frac{\sqrt{\sin^2\theta-\sin^2\eta}}{|\sin\theta-\sin\beta|},
\qquad \theta\in[\eta,\pi-\eta]\cup[\eta-\pi,-\eta],
\]
where the angles $\eta\in[0,\pi/2)$ and $\beta\in(-\pi,\pi]$ are defined
by
\[
\sin\eta=|a|,
\qquad
\sin\beta=-\im a,
\qquad
\text{sign}(\cos\beta)=\text{sign}(\re a).
\]
The choice of the sign of $\cos\beta$ does not affect for the
weight, but will be important for the discussion of the mass points.

Thus the weight is supported on two symmetric arcs of angular amplitude
$2\eta$ centered at $\pm i$.

Concerning the singular part of the measure, its support must lie in
$\{e^{i\theta} : \lim_{r\uparrow1} F(re^{i\theta}) = \infty \}$.
From the expression of $F(z)$ given previously we see that there is
only one possible point in this support: $e^{i\beta}$. Thus it can
only be a mass point with a mass given by (\ref{MASS}), which yields
\[
\mu(\{e^{i\beta}\})=\frac{|\re a|}{\sqrt{1-|\im a|^2}} =
\frac{\sqrt{\sin^2\eta-\sin^2\beta}}{|\cos\beta|}.
\]
The mass point is located outside the support of the weight because
$|\sin\beta|\leq\sin\eta$. Indeed, $\beta$ lies on $(-\eta,\eta)$ or in
its symmetric arc depending whether $\re a>0$ or $\re a<0$. In the limit
case $\re a=0$ the mass point dissapears.


\bigskip

\begin{thebibliography}{99}


\bibitem{A}
A. Ambainis,
\emph{Quantum walks and their algorithmic applications},
International Journal of Quantum Information \textbf{1} (2003) 507--518.


\bibitem{ABNVW}
A. Ambainis, E. Bach, A. Nayak, A. Vishwanath, J. Watrous,
{\em One dimensional quantum walks},
Proc. of the ACM Symposium on Theory and Computation (STOC'01), July 2001, ACM, NY, 2001, pp. 37--49.


\bibitem{BP}
A. Bressler, R. Pemantle,
{\em Quantum random walks in one dimension via generating functions},
DMTCS proc. AH, 2007, pp. 403--414.

\bibitem{BHJ}
O. Bourget, J. S. Howland, A. Joye,
{\em Spectral Analysis of Unitary Band Matrices},
Commun. Math. Phys. \textbf{234} (2003) 191--227.

\bibitem{FIVE}
M. J. Cantero, L. Moral, L. Vel\'azquez,
\emph{Five-diagonal matrices and zeros of orthogonal polynomials on the unit circle},
Linear Algebra Appl. \textbf{362} (2003) 29--56.

\bibitem{MIN}
M.J. Cantero, L. Moral, L. Vel\'azquez,
\emph{Minimal representations of unitary operators and orthogonal polynomials on the unit circle},
Linear Algebra Appl. \textbf{405} (2005) 40--65.

\bibitem{CIR}
H. Carteret, M. Ismail, B. Richmond,
{\em Three routes to the exact aymptotics for the one dimensional random quantum walk},
J. Physics A \textbf{36} (2003) 8775--8795.


\bibitem{DaPuSi08}
D. Damanik, A. Pushnitski, B. Simon,
\emph{The analytic theory of matrix orthogonal polynomials},
Surveys in Approximation Theory \textbf{4} (2008) 1--85.

\bibitem{DRSZ}
H. Dette, B. Reuther, W. Studden, M. Zygmunt,
{\em Matrix measures and random walks with a block tridiagonal transition matrix},
SIAM J. Matrix Anal. Appl. \textbf{29}, No. 1 (2006) 117--142.

\bibitem{F}
W. Feller,
{\em On second order differential operators},
Ann. of Math. \textbf{61}, No. 1 (1955) 90--105.


\bibitem{Ge61}
Ya. L. Geronimus,
\emph{Orthogonal Polynomials},
Consultants Bureau, New York, 1961.


\bibitem{G1}
F. A. Gr\"unbaum,
{\em Random walks and orthogonal polynomials: some challenges},
in Probability, Geometry and Integrable systems, Mark Pinsky and Bjorn Birnir editors,
MSRI publication vol 55, 2007, pp. 241--260, see also arXiv math PR/0703375.

\bibitem{G2}
F. A. Gr\"unbaum,
{\em QBD processes and matrix valued orthogonal polynomials: some new explicit examples},
Dagstuhl Seminar Proceedings 07461, Numerical Methods in structured Markov Chains, 2007, D. Bini editor.

\bibitem{G3}
F. A. Gr\"unbaum,
{\em Block tridiagonal matrices and a beefed-up version of the Ehrenfest urn model},
in M. G. Krein 100's anniversary volume, Odessa, Ukraine, 2008.

\bibitem{GI}
F. A. Gr\"unbaum, M. D. de la Iglesia,
{\em Matrix valued orthogonal polynomials arising from group representation theory and a family of quasi-birth-and-death processes},
SIAM J. Matrix Anal. Appl. \textbf{30}, No. 2 (2008) 741--761.

\bibitem{KMc}
S. Karlin, J. McGregor, {\em Random walks},
Illinois J. Math. \textbf{3} (1959) 66--81.

\bibitem{Ka76}
T. Kato,
{\em Perturbation theory for linear operators},
second edition, Springer-Verlag, Berlin-New York, 1976.

\bibitem{K}
J. Kempe,
{\em Quantum random walks-an introductory overview},
Contemporary Physics \textbf{44}, No. 4 (2003) 307--327.

\bibitem{Ko}
N. Konno,
{\em Quantum walks},
in Quantum Potential Theory, U. Franz, M. Sch\"urmann, editors,
Lecture notes in Mathematics 1954, Springer Verlag, Berlin Heidelberg, 2008.

\bibitem{K1}
M. G. Krein,
{\em  Fundamental aspects of the representation theory of hermitian operators with deficiency index $(m,m)$},
AMS Translations, Series 2, vol. 97, Providence, Rhode Island (1971), pp. 75--143.

\bibitem{K2}
M. G. Krein,
{\em Infinite $J$-matrices and a matrix moment problem},
Dokl. Akad. Nauk SSSR \textbf{69}, No. 2 (1949) 125--128.

\bibitem{LR}
W. Ledermann, G. E. Reuter,
{\em Spectral theory for the differential equations of simple birth and death processes},
Philos. Trans. Roy. Soc. London, Ser. A \textbf{246} (1954) 321--369.

\bibitem{McK}
H. P. McKean, jr.,
{\em Elementary solutions for certain parabolic partial differential equations},
Trans. Amer. Math. Soc. \textbf{82} (1956) 519--548.

\bibitem{M}
D. Meyer,
{\em From quantum cellular automata to quantum lattice gases},
J. Stat. Physics \textbf{85} (1996) 551--574, quant-ph/9604003.

\bibitem{NV}
A. Nayak, A. Vishwanath,
{\em Quantum walk on the line},
Center for Discrete Mathematics $\&$ Theoretical Computer Science, 2000, quant-ph/0010117.



\bibitem{Po}
G. Polya,
{\em  \"Uber eine Aufgabe der Wahrscheinlichkeitsrechnung betreffend die Irrfahrt im Strassennetz},
Mathematische Annalen \textbf{84} (1921) 149--160.

\bibitem{Ru74}
W. Rudin,
{\em Real and Complex Analysis},
second edition, McGraw-Hill Book Co., New York-D\"{u}sseldorf-Johannesburg, 1974.

\bibitem{Si04-1}
B. Simon,
\emph{Orthogonal Polynomials on the Unit Circle, Part 1: Classical Theory},
AMS Colloq. Publ., vol. 54.1, AMS, Providence, RI, 2005.



\bibitem{SiFIVE}
B. Simon,
\emph{CMV matrices: Five years after},
J. Comput. Appl. Math. \textbf{208} (2007) 120--154.

\bibitem{SKJ}
M. Stefanak, T. Kiss, I. Jex,
{\em Recurrence properties of unbiased coined quantum walks on infinite $d$ dimensional lattices},
arXiv: 0805.1322v2 [quant-ph] 4 Sep 2008.

\bibitem{Sz75}
G. Szeg\H o,
\emph{Orthogonal Polynomials},
4th ed., AMS Colloq. Publ., vol. 23, AMS, Providence, RI, 1975.

\bibitem{Wa93}
D. S. Watkins,
\emph{Some perspectives on the eigenvalue problem},
SIAM Rev. \textbf{35} (1993) 430--471.

\end{thebibliography}
\end{document}